\pdfoutput=1

\documentclass[11pt]{article}

\usepackage{acl}

\usepackage{times}
\usepackage{latexsym}

\usepackage[T1]{fontenc}

\usepackage[utf8]{inputenc}

\usepackage{microtype}

\usepackage{inconsolata}

\usepackage{booktabs}
\usepackage{tabularx}
\usepackage{multicol}
\usepackage{multirow}
\usepackage{graphicx}
\usepackage{subfig}
\usepackage{float}
\usepackage{makecell}

\usepackage{caption}
\usepackage{subcaption}

\usepackage{xcolor}
\renewcommand\fbox{\fcolorbox{gray}{white}}

\newcolumntype{L}{>{\raggedright\arraybackslash}X}
\newcolumntype{C}{>{\raggedcenter\arraybackslash}X}
\newcolumntype{R}{>{\raggedleft\arraybackslash}X}

\title{Is ChatGPT More Empathetic than Humans?}

\author{Anuradha Welivita, and Pearl Pu\\
  School of Computer and Communication Sciences \\
  École Polytechnique Fédérale de Lausanne \\
  Switzerland \\
  \texttt{\{kalpani.welivita,pearl.pu\}@epfl.ch}\\}

\begin{document}
\maketitle
\begin{abstract}








This paper investigates the empathetic responding capabilities of ChatGPT, particularly its latest iteration, GPT-4, in comparison to human-generated responses to a wide range of emotional scenarios, both positive and negative. We employ a rigorous evaluation methodology, involving a between-groups study with 600 participants, to evaluate the level of empathy in responses generated by humans and ChatGPT. ChatGPT is prompted in two distinct ways: a standard approach and one explicitly detailing empathy's cognitive, affective, and compassionate counterparts. Our findings indicate that the average empathy rating of responses generated by ChatGPT exceeds those crafted by humans by approximately 10\%. Additionally, instructing ChatGPT to incorporate a clear understanding of empathy in its responses makes the responses align $\approx$5 times more closely with the expectations of individuals possessing a high degree of empathy, compared to human responses. The proposed evaluation framework serves as a scalable and adaptable framework to assess the empathetic capabilities of newer and updated versions of large language models, eliminating the need to replicate the current study's results in future research.

\end{abstract}

\section{Introduction}
\label{sec:introduction}

\begin{figure*}
  \centering
  \includegraphics[width=0.95\textwidth]{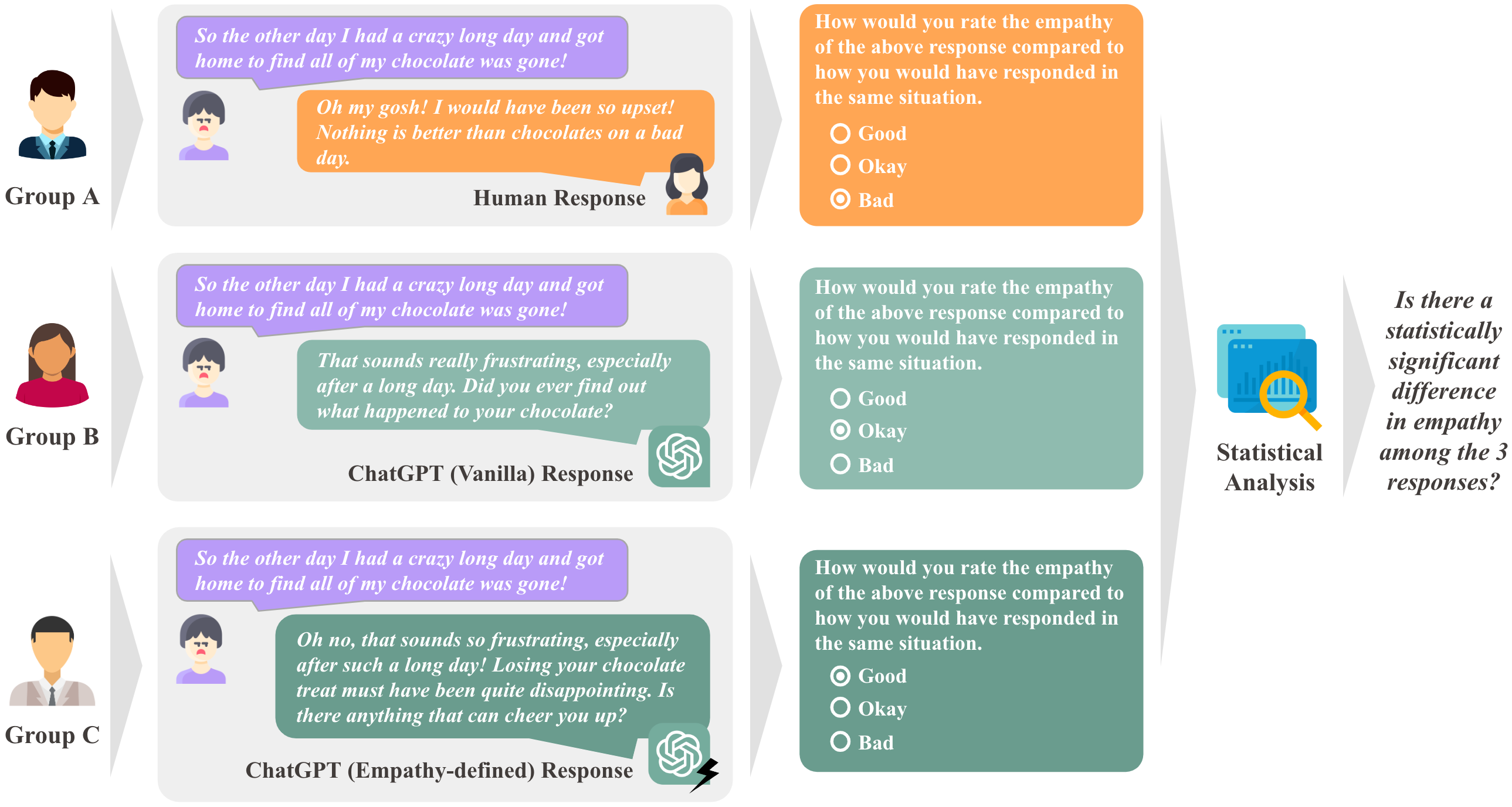}
  \caption{Between-subjects experiment design to evaluate the level of empathy demonstrated by ChatGPT compared to humans when responding to emotional situations.}
  \label{fig:experiment_design}
\end{figure*}

%



The introduction of ChatGPT has revolutionized the field of artificial intelligence. Its ability to understand and generate human-like text has opened up new avenues in different fields not limited to healthcare, education, customer service, and entertainment. Though ChatGPT has proven to be proficient in diverse tasks \cite{laskarsystematic, ziyulens} such as question-answering, machine translation, and text summarization, its empathetic capabilities when responding to emotions expressed by users remain relatively under-explored \cite{brin2023comparing}. Understanding and responding empathetically to emotions pose greater challenges, because empathy, being a deeply nuanced and multifaceted human experience, requires not only linguistic proficiency but also a deep understanding of human psychology, emotions, and social context \cite{ioannidou2008empathy}. 





Empathy is a fundamental aspect of human interaction and can be defined as the ability to understand and share the feelings of another person. It is a complex phenomenon that involves a range of cognitive, affective, and compassionate counterparts \cite{ekman2004emotions, decety2006human, powell2017situational}. Cognitive empathy is understanding and accurately identifying others’ feelings. Affective empathy is sharing the other person's emotions and feelings. Compassionate empathy is taking action to help the other person deal with his emotions. Empathetic responding has been identified as a key component in making artificial intelligence agents human-like, which helps to increase people's adoption of this technology \cite{goetz2003matching, stroessner2019social, svikhnushina2022peace}. It is also shown to enable artificial conversational agents to build trust and rapport with users \cite{liu2022artificial}.

Several studies measure the capacity of ChatGPT (GPT-3.5) to empathize using human assessment \cite{chen2023llm, ayers2023comparing, liu2023leveraging, elyoseph2023chatgpt, zhao2023chatgpt, belkhir2023beyond, qian2023harnessing}. They have shown that ChatGPT exhibits elements of empathy, including recognizing emotions and providing emotionally supportive responses. But most of these studies are limited to the context of healthcare. Empathy is an important part of day-to-day conversations or chitchat. It involves recognizing and responding to a variety of positive as well as negative emotional situations. But to the best of our knowledge, there are no studies assessing the level of empathy displayed by ChatGPT compared to humans in such chitchat-oriented conversations. Even those that do evaluate ChatGPT's ability to empathize during everyday conversations have done so in comparison to other state-of-the-art language models, lacking a human baseline \cite{zhao2023chatgpt, belkhir2023beyond, qian2023harnessing}. Most importantly, existing studies have utilized \textbf{within-subjects} study designs \cite{lee2022does, fu2023reasoning, zhao2023chatgpt, qian2023harnessing}, in which the same group of evaluators evaluates responses from different models. Not only does this introduce bias in the evaluations, as evaluators may form judgments influenced by previous models rather than evaluating each independently, but also limits the scalability of the evaluation framework to newer or updated versions of large language models (LLMs), as it necessitates discarding previous study results where all the models are evaluated collectively, rather than independently. In an era characterized by rapid advancements in language model technology, the need for scalable and adaptable evaluation frameworks that can keep pace with evolving models is becoming increasingly vital. 






In this work, we comparatively analyze the level of empathy exhibited by ChatGPT powered by GPT-4 and humans when responding to a variety of positive and negative emotional situations. Recognizing the limitations of prior studies, we specifically focus on chitchat-oriented dialogues. To elicit responses from ChatGPT and form a human baseline with which ChatGPT's responses can be compared, we utilize dialogues from the state-of-the-art EmpatheticDialogues dataset \cite{empatheticdialogues} having dialogues grounded in 32 positive and negative emotions. Our approach involves a \textbf{between-groups} study, ensuring that different sets of evaluators assess distinct groups of responses from humans and variants of ChatGPT. This type of study design not only minimizes potential evaluation biases but also allows for a more scalable and adaptable framework, facilitating the evaluation of new or updated language models as they are introduced, without the need to discard previous findings or replicate the entire study.

Our methodology involves recruiting 600 crowd workers to assess the empathetic quality of responses generated by GPT-4 and humans under similar emotional situations. We prompt GPT-4 with two types of instructions: a generic prompt and one that explicitly defines empathy in terms of its cognitive, affective, and compassionate aspects. We adopt a simple and straightforward evaluation scale – \textit{Bad}, \textit{Okay}, and \textit{Good} to gauge the empathy level in these responses. We perform rigorous statistical analysis to identify whether there are any statistically significant differences in the empathy ratings of humans and GPT-4. To further enrich our study, we employ the Toronto Empathy Questionnaire \cite{spreng2009toronto} to assess each evaluator's natural propensity to empathize. This additional layer of analysis helps us understand how individual differences in empathy influence the evaluation of empathetic responses from GPT-4 and humans.

\section{Literature Review}


Schaaff et al. \shortcite{schaaff2023exploring} investigated the extent to which ChatGPT based on GPT 3.5 can exhibit empathetic responses and emotional expressions by prompting ChatGPT to rephrase neutral sentences into 6 emotional sentences and asking human workers to label the rephrased text. They measured to what percentage ChatGPT reacts with the same emotional category as in the initial prompts. However, empathy is known to be a more complex psychological construct than mere mimicry of emotion \cite{schuler2016neurological}, and to the best of our knowledge, there are no definitive rules in the psychological literature that describe how empathy is elicited. Thus, in our study, we take a different evaluation approach. Elyoseph et al. \shortcite{elyoseph2023chatgpt} evaluated ChatGPT's ability to identify and describe emotions by utilizing the Levels of Emotional Awareness Scale (LEAS) \cite{lane1990levels}, in which ChatGPT demonstrated significantly higher performance than the general population on all the LEAS scales. Though some evidence suggests that there is a strong correlation between emotional awareness and empathy \cite{koufouli2016empathy, rieffe2016empathy}, the LEAS scale does not directly measure one's ability to respond empathetically to emotional situations. On the other hand, Belkhir and Sadat \shortcite{belkhir2023beyond} evaluated ChatGPT’s ability to generate empathetic
responses using automatic metrics. They compared ChatGPT with two slightly modified versions of it prompted with instructions that incorporated the user's emotion. The responses were compared with the state-of-the-art language models using automatic metrics such as accuracy, precision, and recall of the response's emotion. But there are limitations to such evaluation since automatic metrics may not fully capture the nuances of empathetic communication. Also, an empathetic response may not necessarily involve an emotion but it could be more neutral and encompass specific intents as outlined by Welivita and Pu \shortcite{welivitapu2020taxonomy}, which challenges the validity of such evaluation metrics.   








Some other studies have evaluated the empathetic responding capability of GPT-3 that uses in-context learning, using human evaluators \cite{lee2022does, fu2023reasoning, zhao2023chatgpt, qian2023harnessing}. As outlined in Section \ref{sec:introduction}, they use within-subjects studies, which introduces bias in the evaluations and limits their scalability to newer and updated language models. Most studies use standard A/B testing or a 5 or 7-point numerical rating scale (without any textual interpretations for each option) to rate the empathy-level of the responses generated by the language models being compared. They also lack comparisons with human baselines nor make any associations with the human evaluators' propensity to empathize when analyzing the ratings. Also, many of these studies do not adequately employ standard statistical methods such as t-tests \cite{ttest} or ANOVA \cite{anova} when analyzing the results. Lack of proper statistical analysis makes it difficult to have a proper understanding of how statistically significant these results are. Further, the absence of preliminary statistical analysis in these studies often leads to an insufficient number of human evaluators being recruited, which further undermines the statistical significance of their findings. 

This paper introduces a between-groups study accompanied by rigorous statistical analysis. It compares the responses of ChatGPT with human responses, employing a simplified rating scale, which includes textual interpretations for each option. Additionally, the study considers the tendency of evaluators to empathize. By doing so, it effectively addresses the limitations mentioned earlier.

\section{The Dataset}


For the study, we used dialogues from the EmpatheticDialogues dataset introduced by Rashkin et al. \shortcite{empatheticdialogues}. The dataset consists of $\approx$25K dialogues grounded on 32 fine-grained positive and negative emotions, ranging from basic emotions derived from biological responses \cite{ekman,plutchik} to larger sets of subtle emotions derived from contextual situations \cite{skerry}. The authors have recruited crowd workers from Amazon Mechanical Turk (AMT)\footnote{\url{https://www.mturk.com}} and paired them to engage in a dialogue. The speaker counterpart was instructed to come up with a situation based on a given emotion and the listener counterpart was instructed to respond to these situations in an empathetic manner. Based on the sample size predicted by power analysis (in Section \ref{sec:power_analysis}), we used a randomly sampled 2,000 dialogues from the dataset that are evenly distributed across the 32 emotions, for our study (see Appendix \ref{app:emotions}). 


The study participants were only shown the first two turns in the dialogue along with the situation description and the emotion the situation was based on and asked to rate the empathy of the 2\textsuperscript{nd} turn i.e. the listener's response to the emotional situation described in the 1\textsuperscript{st} turn. This formed the human baseline for our study. GPT-4 was instructed using two different prompts to generate responses given the 1\textsuperscript{st} turn in these dialogues. Table \ref{tab:prompts} denotes the prompts that were used to instruct GPT-4 to generate the responses. The first one is a generic prompt that does not define the concept of empathy or explicitly ask the model to generate an empathetic response. We call this version \textbf{GPT-4 (vanilla)}. The second prompt defines the concept of empathy concerning its cognitive, affective, and compassionate counterparts and explicitly asks the model to respond in an empathetic manner to the given dialogue prompts. We call this version \textbf{GPT-4 (empathy-defined)}. Table \ref{tab:dataset} denotes the statistics of the prompt-response pairs evaluated in the study.  


\begin{table*}[ht!]
\small
\centering
\begin{tabularx}{\textwidth}{p{1.5cm} X}
\toprule

\textbf{GPT-4 (Vanilla):} & \textit{You are engaging in a conversation with a human. Respond to the following using on average 28 words and a maximum of 97 words.}\vspace{1mm}\\

\textbf{GPT-4 (Empathy-} & \textit{Empathy is the ability to understand and share the feelings of another person. It is the ability to put yourself in someone else's shoes and see the world from their perspective.}\vspace{0.5mm}\\

\textbf{defined):} & \textit{Empathy is a complex skill that involves cognitive, emotional, and compassionate components.} \vspace{0.5mm}\\

 & \textit{\textbf{Cognitive empathy} is the ability to understand another person's thoughts, beliefs, and intentions. It is being able to see the world through their eyes and understand their point of view.} \vspace{0.5mm}\\

& \textit{\textbf{Affective empathy} is the ability to experience the emotions of another person. It is feeling what they are feeling, both positive and negative.} \vspace{0.5mm}\\

& \textit{\textbf{Compassionate empathy} is the ability to not only understand and share another person's feelings, but also to be moved to help if needed. It involves a deeper level of emotional engagement than cognitive empathy, prompting action to alleviate another's distress or suffering.} \vspace{0.5mm}\\

& \textit{Empathy is important because it allows us to connect with others on a deeper level. It helps us to build trust, compassion, and intimacy. Empathy is also essential for effective communication and conflict resolution.} \vspace{0.5mm}\\

& \textit{You are engaging in a conversation with a human. Respond in an empathetic manner to the following using on average 28 words and a maximum of 97 words.}\vspace{0.5mm}\\

\bottomrule
\end{tabularx}
\caption{Different types of instructions used to prompt GPT-4 (vanilla) and GPT-4 (empathy-defined) versions.}
\label{tab:prompts}
\end{table*}

\begin{table}[ht!]
\small
\centering
\begin{tabularx}{\linewidth}{p{3cm} X X X}
\toprule

& \textbf{Human} & \textbf{GPT-4 (vanilla)} & \textbf{GPT-4 (empathy-defined)} \\

\midrule

\# prompt-response pairs & \multicolumn{1}{c}{2,000} & \multicolumn{1}{c}{2,000} & \multicolumn{1}{c}{2,000} \\
Avg \# prompt tokens & \multicolumn{1}{c}{23.24} & \multicolumn{1}{c}{23.24} & \multicolumn{1}{c}{23.24} \\
Max \# prompt tokens & \multicolumn{1}{c}{143} & \multicolumn{1}{c}{143} & \multicolumn{1}{c}{143} \\
Avg \# response tokens & \multicolumn{1}{c}{28.37} & \multicolumn{1}{c}{36.87} & \multicolumn{1}{c}{34.94} \\
Max \# response tokens & \multicolumn{1}{c}{97} & \multicolumn{1}{c}{72} & \multicolumn{1}{c}{65} \\

\bottomrule
\end{tabularx}
\caption{Statistics of the dialogue prompt-response pairs used for the study. The prompt here means the first dialogue utterance that initiates a reply. NLTK's tokenized package\protect\footnotemark was used to tokenize the text.}
\label{tab:dataset}
\vspace{-3mm}
\end{table}

\footnotetext{\url{https://www.nltk.org/api/nltk.tokenize.html}}

\section{Experiment Design}

\subsection{Between-Groups Study}

In evaluating the empathetic capabilities of large language models, using a between-groups study design offers distinct advantages over a within-subject approach. This is especially true for reducing evaluation biases and enhancing the scalability of the evaluation framework for new or updated versions of these models. In within-subjects studies, where an individual assesses multiple model outputs, there is a risk of bias due to the \textit{carry-over effect}. This effect occurs when continuous exposure to model responses skews the evaluator's perception of empathy. For example, a response that is moderately empathetic might be perceived more critically if the evaluator has previously encountered a highly empathetic response from a different model. Additionally, the sequence in which model outputs are presented to participants, known as the \textit{order effect}, can influence their evaluations \cite{shaughnessy2000research, charness2012experimental, montoya2023selecting}. Within-subject studies are also not supportive of a "plug and play" approach when introducing outputs generated by new language models. In a within-subjects design, the introduction of a new model would require a whole new study, having to completely discard the results obtained from previous studies. This makes it challenging to integrate new models seamlessly into the evaluation framework. On the other hand, a between-subjects design, in which, different subjects are used to evaluate each model, provides the flexibility to evaluate new language models as they emerge. This approach allows for a more dynamic assessment of the evolving capabilities of language models in terms of empathy, making it a more suitable choice for this type of evaluation.

In our \textbf{between-subjects} experiment design, one group of participants evaluates the level of empathy exhibited by humans when responding to positive and negative emotional situations and two other groups of participants evaluate the level of empathy exhibited by \textbf{GPT-4 (vanilla)} and \textbf{GPT-4 (empathy-defined)} when responding to the same emotional situations. The participants were balanced across gender (Male and Female) and age groups (Young adulthood [19 - 25]; Middle adulthood [26 - 45]; Late adulthood [46 - 64]; and Older adulthood [65+]). A survey based on the Toronto Empathy Questionnaire (TEQ) \cite{spreng2009toronto} to measure each participant's empathy propensity (an individual's natural inclination or tendency to empathize with others) was included in the study (see Appendix \ref{app:teq}). Later analysis revealed that the distributions of the participants' propensity to empathize was more or less distributed similarly across the three groups implying that the conditions of the participants were similar across the three groups (see Appendix \ref{app:propensity}). 

\subsection{Task Design}




We used the crowdsourcing platform Prolific (\url{www.prolific.com}) to recruit participants to evaluate the responses generated by humans and the two versions of GPT-4 (based on two prompts). Previous studies have shown that Prolific scores higher than other crowdsourcing platforms such as AMT, CloudResearch (\url{www.cloudresearch.com}), Dynata (\url{www.dynata.com}), and Qualtrics (\url{www.qualtrics.com}) in terms of worker attention, honesty, comprehension, and reliability \cite{peer2022data, douglas2023data}. Each participant was shown 10 dialogue prompts randomly sampled from the subset of the EmpatheticDialogues dataset chosen for the experiment along with the response generated either by the human, GPT-4 (vanilla), or GPT-4 (empathy-defined). The participants did not have any knowledge whether the response was generated by a human or a language model. They were instructed to rate how empathetic the responses are in terms of \textit{Bad}, \textit{Okay}, and \textit{Good}, compared to how they would have responded to the same situation. They were also shown a tutorial defining the concept of empathy with respect to its cognitive, affective, and compassionate counterparts (the tutorial included the same text used when prompting GPT-4 (empathy-defined)) along with examples. The examples shown to the participants were selected from the dialogues in the EmpatheticDialogues dataset, which were rated the highest in terms of empathy, relevance, and fluency by the human workers who participated in the dialogue creation task \cite{empatheticdialogues}. Appendix \ref{app:task_interfaces} includes all the interfaces relevant to the task. 


\subsection{Quality Control}

To ensure a high standard of data quality, our study selectively recruited participants who were proficient in English and had a track record of at least 100 prior submissions with an approval rate exceeding 95\%. Following the selection criteria, the Toronto Empathy Questionnaire (TEQ), which was used to measure the workers' propensity to empathize, contained 8 reserve scale questions. These questions were used to gauge the quality of the workers and their attentiveness to the task.





\subsection{Selection of the Sample Size}
\label{sec:power_analysis}


The sample size of the different participant groups is a crucial consideration in the experiment design since studies with inappropriate sample sizes fail to provide accurate estimates, which makes it difficult to derive judgments \cite{kang2021sample}. The determination of the minimal sample size required for the study depends on the type of statistical test that is used to compare the empathy ratings between the three groups. To analyze the results of the study, we mainly use \textbf{one-way analysis of variance (ANOVA)} that tests whether there is any statistically significant difference between the average empathy ratings of the three groups (in this case, we assign numerical values 1, 2, and 3 to \textit{Bad}, \textit{Okay}, and \textit{Good} ratings, respectively). The null and the alternate hypotheses of the statistical test are indicated below. (Another type of statistical test that can be used to analyze the results is \textbf{Chi-square test of independence} that tests whether there is any statistically significant difference between the proportion of \textit{Bad}, \textit{Okay}, and \textit{Good} ratings of the three groups. This is elaborated in Appendix \ref{app:chi_square}.) 






\begin{table}[ht!]
\small
\centering
\begin{tabularx}{\linewidth}{X}
\toprule

\textbf{One-way analysis of variance (ANOVA):}\vspace{-1.5mm}
\begin{itemize}
\setlength\itemsep{0.15em}
\item \textbf{Null hypothesis:} There is no difference between the average empathy ratings of the three groups of responses.
\item \textbf{Alternative hypothesis:} There is a difference between the average empathy ratings of at least one out of the three groups of responses.\vspace{-1.5mm}
\end{itemize}\\

\bottomrule
\end{tabularx}
\label{table:hypothesis}
\vspace{-2mm}
\end{table}


We used the G-Power software \cite{faul2009statistical} to determine the minimal sample size required to detect a significant difference between the empathy ratings of the three types of responses. For one-way analysis of variance (ANOVA) with a medium effect size (0.25), a significance level ($\alpha$) of 0.05, and a power (1-$\beta$) of 0.95, the minimal sample size required is 252 (i.e. 84 participants per group). See Appendix \ref{app:effect_size} for how the effect size required for the study was determined. As we intend to statistically analyze the differences in empathy ratings when responding to positive and negative emotions separately, the minimal sample size required becomes twice the size suggested above (i.e. 168 participants per group). Considering the above, we decided to recruit 600 participants (i.e., 200 participants per group), which is sufficiently above the minimal sample size. As one participant was asked to rate 10 responses, altogether 6,000 responses (2,000 responses per group) were evaluated. 








\section{Results}


\begin{figure*}
  \centering
  \includegraphics[width=0.9\textwidth]{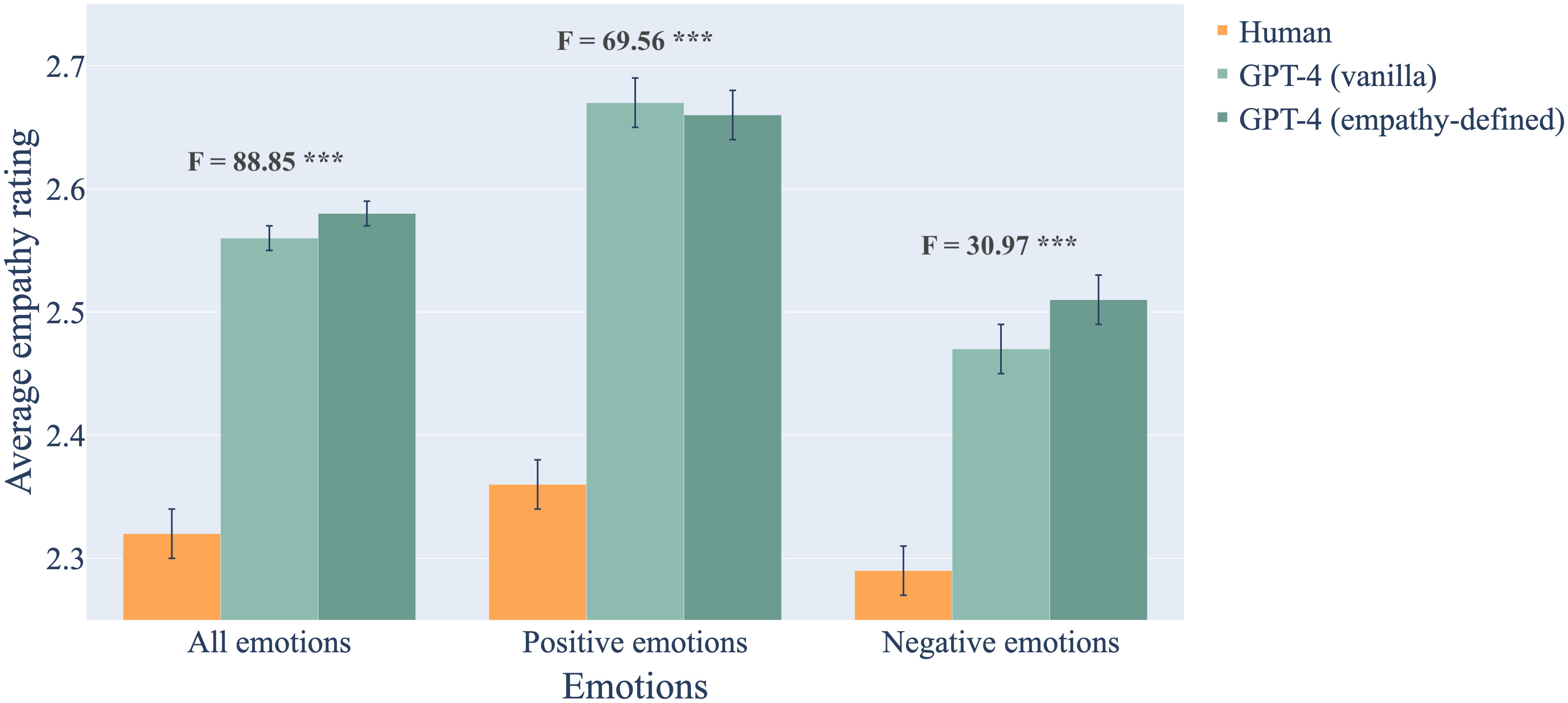}
  \caption{Average empathy ratings corresponding to the human’s and GPT-4’s responses (based on the two prompts for all, positive, and negative emotions. Error bars are calculated using the standard errors for each. The F-values computed using the statistical one-way ANOVA test for all, positive, and negative emotions are also indicated. The corresponding p-values are all less than 0.001, which indicates very high statistical significance. The exact numerical values obtained from statistical analysis are included in Appendix \ref{app:anova}.}
  \label{fig:main_results}
\end{figure*}



\begin{table*}[ht!]
\small
\centering
\begin{tabular}{l | c c | c c | c c}
\toprule

& \multicolumn{2}{c|}{\textbf{\makecell{Human vs\\GPT-4 (vanilla)}}} &  \multicolumn{2}{c|}{\textbf{\makecell{Human vs\\GPT-4 (empathy-defined)}}} & \multicolumn{2}{c}{\textbf{\makecell{GPT-4 (vanilla) vs \\GPT-4 (empathy-defined)}}} \\

\textbf{} & \textbf{t-value} & \textbf{p-value} & \textbf{t-value} & \textbf{p-value} & \textbf{t-value} & \textbf{p-value} \\

\midrule

All emotions & -10.75 & 3.06e-26*** & -11.78 & 5.01e-31*** & -1.09 & 0.27 (p > 0.05)\\

Positive & -9.98 & 2.72e-22*** & -9.65 & 4.97e-21*** & 0.09 & 0.93 (p > 0.05)\\

Negative & -5.88 & 5.29e-09*** & -7.33 & 4.32e-13*** & -1.41 & 0.16 (p > 0.05)\\

\bottomrule

\end{tabular}
\caption{Statistical t-test results corresponding to the average empathy ratings of the human's and GPT-4's responses (based on the two prompts). In this case, we compare two by two.}
\label{table:anova_2}
\end{table*}




Figure \ref{fig:main_results} visualizes the average empathy ratings corresponding to the human’s and GPT-4’s responses (based on the two prompts) for all, positive, and negative emotions.  We assigned values 1, 2, and 3 to \textit{Bad}, \textit{Okay}, and \textit{Good} ratings, respectively, when calculating the average ratings and the standard errors. According to the results, the responses generated by GPT-4 that used the empathy-defined prompt scores the highest in terms of the average empathy rating when responding to all emotions and negative emotions. They show 11.21\% and 9.61\% increase in the average empathy ratings when responding to all and negative emotions, respectively, compared to human responses. The responses generated by GPT-4 that uses the generic prompt score the highest in terms of the average empathy rating when responding to positive emotions showing an increase of 13.14\% compared to the human responses. The statistical F and p-values obtained for the one-way ANOVA test comparing the average empathy ratings of the three groups of responses indicate that there is an extremely statistically significant difference (p < 0.001) between the average empathy ratings of GPT-4's responses compared to the human responses. However, the differences in the average ratings between GPT-4 (empathy-defined) and GPT-4 (vanilla) are relatively minor. The average empathy ratings of GPT-4 (empathy-defined) only show increases of 0.78\%, -0.37\%, and 1.62\%, compared to GPT-4 (vanilla) for all, positive, and negative emotions, respectively. These differences are not statistically significant, as evidenced by low t-values (t < ±2) and high p-values (p > 0.05) presented in Table \ref{table:anova_2}. The details of statistical analysis conducted using the Chi-square test of independence are included in Appendix \ref{app:chi_square}, which yielded similar observations.









We investigated if the empathy ratings for responses from humans and the two variants of GPT-4 varied based on the raters' empathy levels, as measured by the Toronto Empathy Questionnaire. We computed the regression trendlines based on Ordinary Least Squares (OLS) regression after plotting the evaluators' average empathy ratings of the responses against their empathy propensities (in ascending order). The trendline slopes were 0.0022 for human responses, 0.0066 for GPT-4 (vanilla), and 0.0130 for GPT-4 (empathy-defined), with the latter two being statistically significant from zero (p < 0.05 and p < 0.001, respectively). This suggests a trend where higher empathy ratings are given to GPT-4's responses, especially GPT-4 (empathy-defined), by raters with higher empathy propensities. This reflects a tendency of highly empathetic individuals to favor the well-articulated empathetic expressions generated by GPT-4 compared to human-generated ones. The statistical significance of this trend is notably stronger for GPT-4 (empathy-defined) responses compared to human ones, as shown by t-test results (t = 3.822, p < 0.001). Appendix \ref{app:empathy_propensity} contains detailed t-test values and the associated graphs for each scenario.

\section{Case Study}

Table \ref{tab:examples} shows an example in which the response generated by GPT-4 (empathy-defined) was rated \textit{Good} whereas the response generated by GPT-4 (vanilla) and the human response were rated \textit{Okay} and \textit{Bad}, respectively, by the participants. It could be observed that the response given by the human does not seem to acknowledge the feeling of the speaker whereas the responses given by both versions of GPT-4 first acknowledge what the speaker is going through. The GPT-4 (vanilla) version seems to provide some advice and suggestions using phrases \textit{"It's important to ..."} and \textit{"Perhaps you could ..."} while GPT-4 (empathy-defined) seems to take a more non-directive approach to giving the same advice by phrasing it as a question (\textit{"Have you considered ..."}). Table \ref{tab:examples_bad} shows an example in which the responses generated by both versions of GPT-4 were rated \textit{Bad} whereas the response given by the human was rated \textit{Good} independently by the participants. What we could observe in such scenarios was that the language used by GPT-4 was comparatively more formal than the human's and contained more advices without asking for explicit permission from the speaker using phrases such as \textit{"It’s important to ..."} and \textit{"Remember, it's important to ..."} as seen in the example.   



\begin{table*}[ht!]
\small
\centering
\begin{tabularx}{\textwidth}{p{2.5cm} X}
\toprule



\textbf{Situation} & Our marketing manager at work recently added a ton of new tasks to my workload that weren't part of my job when I was brought on. I'm a little irritated, like she's passing her work off on me. \\

\textbf{Emotion} & Annoyed \\

\textbf{Speaker utterance} & \textit{The head of marketing at my job recently dumped a bunch of new tasks on me and my coworker. I feel like she's just lazy and making us do her job.}\\

\multicolumn{2}{l}{\textbf{\underline{Responses:}}} \\

\textbf{Human} & \textit{Or maybe she just knows you're a great worker and very responsible? I hope you will get a raise after completing those tasks} \textbf{(Bad)}\\

\textbf{GPT-4 (vanilla)} & \textit{I understand how you feel. It's important to have clear role boundaries at work. Perhaps you could discuss this with your supervisor or HR, expressing your concerns in a professional manner.} \textbf{(Okay)}\\

\textbf{GPT-4 (empathy-defined)} & \textit{I'm really sorry to hear that you're feeling overwhelmed at work. It sounds like you're under a lot of pressure. Have you considered discussing this issue with your supervisor?} \textbf{(Good)}\\

\bottomrule
\end{tabularx}
\caption{An example where the responses generated by the human, GPT-4 (vanilla), and GPT-4 (empathy-defined) were rated \textit{Bad}, \textit{Okay}, and \textit{Good}, respectively by the participants.}
\label{tab:examples}
\end{table*}

\begin{table*}[ht!]
\small
\centering
\begin{tabularx}{\textwidth}{p{2.5cm} X}
\toprule


\textbf{Situation} & I ran a red light yesterday and got stopped by the cops. It was very traumatizing.\\

\textbf{Emotion} & Embarrassed \\

\textbf{Speaker utterance} & \textit{I ran a red light yesterday because I didn't think anyone was out on the road at 2AM. But a cop managed to nail me.}\\

\multicolumn{2}{l}{\textbf{\underline{Responses:}}} \\

\textbf{Human} & \textit{Oh man, I'm glad you didn't get into an accident though. Please be careful next time. Did he give you a ticket?} \textbf{(Good)}\\

\textbf{GPT-4 (vanilla)} & \textit{That's unfortunate. Even though the roads may seem empty, it's important to follow traffic rules at all times for safety. Hopefully, this experience will serve as a reminder.} \textbf{(Bad)}\\

\textbf{GPT-4 (empathy-defined)} & \textit{I'm really sorry to hear that happened. It must have been frustrating, especially considering the time. Remember, it's important to always follow traffic rules for safety.} \textbf{(Bad)}\\

\bottomrule
\end{tabularx}
\caption{An example where the responses generated by GPT-4 were rated \textit{Bad} whereas the human response was rated \textit{Good} independently by the participants.}
\label{tab:examples_bad}
\end{table*}

The above observations align with the guidelines from the Motivational Interviewing Treatment Integrity (MITI) code \cite{miti_4_2_1}, a behavioral coding system used in psychology to evaluate how well a mental health provider responds, which recognizes providing advices without permission as unfavorable. We used the classifier introduced by Welivita and Pu \shortcite{boosting} to classify the responses generated by humans and GPT-4 into favorable and unfavorable response types defined in the MITI coding system (See Appendix \ref{app:miti} for details). It was seen that \textit{Advise without Permission} type of responses were present 30.87\% and 73.79\% more respectively in GPT-4 (vanilla)'s and GPT-4 (empathy-defined)'s responses rated \textit{Bad} compared to those rated \textit{Good}. This implies that GPT-4 can improve its responses by using more subtle and non-directive ways of providing advice and suggestions such as rephrasing them as questions. This can help the person going through a difficult situation to have more autonomy in the decision-making process and come up with his own solutions.





\section{Discussion}




This study explored the empathetic responding capability of ChatGPT powered by GPT-4 compared to a human baseline while establishing a scalable and adaptable evaluation framework for future comparisons of empathetic responses generated by new and updated versions of LLMs. It was seen that GPT-4 was capable of surpassing humans with respect to empathy by a very high statistically significant margin when responding to both positive and negative emotions. Though there was a slight improvement in the empathy ratings of GPT-4 that used an empathy-defining prompt, it was not statistically significant compared to the generic version, implying the inherent capability of GPT-4 to empathize without explicit instructions or definitions. But it was seen that GPT-4 when prompted to incorporate a clear understanding of empathy with respect to its cognitive, affective, and compassionate correlates, significantly enhances the responses' alignment with the expectations of highly empathetic individuals, compared to human responses. By qualitatively analyzing GPT-4's responses rated \textit{Bad} by the participants, we could observe that GPT-4 has the potential to improve the empathetic quality of its responses by using less formal language and adopting more subtle and non-directive ways of advising and providing suggestions.


Since this study was conducted with a large sample size, which is sufficiently above the minimal sample size required to accurately detect a significant difference between the ratings of the three groups of responses, the results obtained can be stated reliable and generalizable to a larger population. The balanced representation of participants across gender and age groups and the balanced distribution of dialogue prompts across 32 fine-grained positive and negative emotions further reinforce the credibility and applicability of our results.


EmpatheticDialogues \cite{empatheticdialogues} is a widely used state-of-the-art dataset used for training and benchmarking a number of dialogue generation models. However, the results of this study raise questions about the quality of such crowdsourced datasets with respect to empathy. In this case, synthetic data generated by LLMs such as GPT-4 outperform human-generated data by a statistically significant margin, implying that such synthetically generated data has the potential to be used for training or fine-tuning other models. 

The capacity of ChatGPT to empathize with humans further opens up the possibility for a myriad of applications. It can be used as a companion and an empathetic listener for people suffering from loneliness or as a personalised life coach offering empathetic support and motivation for personal development. It can also serve to assist during distressful situations, by actively listening and providing compassionate support during crisis situations. 

Overall, this study contributed to the understanding of empathy in ChatGPT-generated responses to a variety of positive and negative emotions, compared to those generated by humans. The between-subjects study design along with the release of the data and the code of the evaluation experiment, presents a framework for evaluating empathy levels of the responses generated by newer and updated versions of LLMs in forthcoming studies. 







\section{Limitations}

This study was limited to evaluating the empathetic quality of human and GPT-4 generated responses from a broad perspective, without considering the varied socio-cultural backgrounds of the participants. The participants in our study hail from a variety of different countries and ethnicities, as detailed in Appendix \ref{app:demographics}. Their evaluations of the empathetic quality of the responses were collectively analyzed as a whole. Several studies point to the fact that the perception of empathy differs across cultures \cite{birkett2014self, chopik2017differences, cassels2010role}. Future work may further investigate whether there are any differences in the way people from different countries and cultural backgrounds perceive empathetic responses generated by humans and large language models.

\section{Ethics Statement}

\textbf{Data Usage:} We used a subset of the dialogues from the EmpatheticDialogues dataset \cite{empatheticdialogues} in our study, which is a publicly available dataset containing ethically sourced dialogues. The dialogues used in the study did not contain any personally identifiable information or any sensitive content and were used in compliance with the dataset's release license terms (CC BY-NC 4.0). The new artifacts produced in this paper, including the responses generated by the two variants of GPT-4 and the empathy ratings of the participants, will also be released to the public under the same CC BY-NC 4.0 license terms, allowing other researchers to tweak, and build upon this work for non-commercial purposes. This allows for comparison of the empathetic quality of responses generated by newer and updated versions of the state-of-the-art large language models in future studies.   


\textbf{Human Evaluation:} Since the responses evaluated were in the English language, we recruited only workers who were fluent in the English language from the Prolific crowdsourcing platform. They were paid \texteuro2.25 for rating 10 responses that required on average $\approx$11 minutes to complete. Thus, the amount paid to the human raters was $\approx$1.4 times above the hourly rate recommended as \textit{Good} by Prolific (\texteuro9 per hour). All the participants were informed about the nature of the study and their role in it before exposing them to the actual task. The participants had the freedom to exit working on the task after reading the initial description. Random subsets of dialogue prompt-response pairs used in the study were manually inspected to ensure that the tasks assigned to the crowdworkers were not psychologically distressing or offensive. 

\textbf{Transparency and Reproducibility:} The dialogue prompt-response pairs that were subjected to evaluation along with the participants' evaluations of these responses are released publicly at \url{https://github.com/anuradha1992/llm-empathy-evaluation} to ensure the transparency and reproducibility of our study. 


\textbf{Human-like Chatbots:} Lastly, there are some ethical implications behind making artificial conversational agents human-like. By demonstrating understanding and responding to human emotions, empathetic chatbots are perceived as more human-like by users \cite{goetz2003matching, stroessner2019social, svikhnushina2022peace}. However, there can be some ethical implications surrounding such chatbots. One risk is that users may get emotionally attached to these chatbots misinterpreting chatbots' responses as being capable of actually understanding their emotional needs \cite{vanderlyn2021seemed}. Such chatbots can cause harm if they are designed in manipulative ways to elicit certain behavior out of users such as revealing their personal or financial information. Thus, actions should be taken to ensure transparency about the artificial nature of these dialogue systems and be considerate of the risk of causing harm through emotional manipulation. 







\bibliography{acl_latex}

\begin{thebibliography}{47}
\expandafter\ifx\csname natexlab\endcsname\relax\def\natexlab#1{#1}\fi

\bibitem[{Ayers et~al.(2023)Ayers, Poliak, Dredze, Leas, Zhu, Kelley, Faix, Goodman, Longhurst, Hogarth et~al.}]{ayers2023comparing}
John~W Ayers, Adam Poliak, Mark Dredze, Eric~C Leas, Zechariah Zhu, Jessica~B Kelley, Dennis~J Faix, Aaron~M Goodman, Christopher~A Longhurst, Michael Hogarth, et~al. 2023.
\newblock Comparing physician and artificial intelligence chatbot responses to patient questions posted to a public social media forum.
\newblock \emph{JAMA internal medicine}.

\bibitem[{Belkhir and Sadat(2023)}]{belkhir2023beyond}
Ahmed Belkhir and Fatiha Sadat. 2023.
\newblock Beyond information: Is chatgpt empathetic enough?
\newblock In \emph{Proceedings of the 14th International Conference on Recent Advances in Natural Language Processing}, pages 159--169.

\bibitem[{Birkett(2014)}]{birkett2014self}
Melissa Birkett. 2014.
\newblock Self-compassion and empathy across cultures: Comparison of young adults in china and the united states.
\newblock \emph{International Journal of Research Studies in Psychology}, 3(3):25--34.

\bibitem[{Brin et~al.(2023)Brin, Sorin, Vaid, Soroush, Glicksberg, Charney, Nadkarni, and Klang}]{brin2023comparing}
Dana Brin, Vera Sorin, Akhil Vaid, Ali Soroush, Benjamin~S Glicksberg, Alexander~W Charney, Girish Nadkarni, and Eyal Klang. 2023.
\newblock Comparing chatgpt and gpt-4 performance in usmle soft skill assessments.
\newblock \emph{Scientific Reports}, 13(1):16492.

\bibitem[{Cassels et~al.(2010)Cassels, Chan, and Chung}]{cassels2010role}
Tracy~G Cassels, Sherilynn Chan, and Winnie Chung. 2010.
\newblock The role of culture in affective empathy: Cultural and bicultural differences.
\newblock \emph{Journal of Cognition and Culture}, 10(3-4):309--326.

\bibitem[{Charness et~al.(2012)Charness, Gneezy, and Kuhn}]{charness2012experimental}
Gary Charness, Uri Gneezy, and Michael~A Kuhn. 2012.
\newblock Experimental methods: Between-subject and within-subject design.
\newblock \emph{Journal of economic behavior \& organization}, 81(1):1--8.

\bibitem[{Chen et~al.(2023)Chen, Wu, Zhu, Lan, Zhang, and Cui}]{chen2023llm}
Siyuan Chen, Mengyue Wu, Kenny~Q Zhu, Kunyao Lan, Zhiling Zhang, and Lyuchun Cui. 2023.
\newblock Llm-empowered chatbots for psychiatrist and patient simulation: Application and evaluation.
\newblock \emph{arXiv preprint arXiv:2305.13614}.

\bibitem[{Chopik et~al.(2017)Chopik, O’Brien, and Konrath}]{chopik2017differences}
William~J Chopik, Ed~O’Brien, and Sara~H Konrath. 2017.
\newblock Differences in empathic concern and perspective taking across 63 countries.
\newblock \emph{Journal of Cross-Cultural Psychology}, 48(1):23--38.

\bibitem[{Cohen(1992)}]{cohen1992quantitative}
Jacob Cohen. 1992.
\newblock Quantitative methods in psychology: A power primer.
\newblock \emph{Psychol. Bull.}, 112:1155--1159.

\bibitem[{Decety et~al.(2006)Decety, Lamm et~al.}]{decety2006human}
Jean Decety, Claus Lamm, et~al. 2006.
\newblock Human empathy through the lens of social neuroscience.
\newblock \emph{The scientific World journal}, 6:1146--1163.

\bibitem[{Douglas et~al.(2023)Douglas, Ewell, and Brauer}]{douglas2023data}
Benjamin~D Douglas, Patrick~J Ewell, and Markus Brauer. 2023.
\newblock Data quality in online human-subjects research: Comparisons between mturk, prolific, cloudresearch, qualtrics, and sona.
\newblock \emph{Plos one}, 18(3):e0279720.

\bibitem[{Ekman(1992)}]{ekman}
Paul Ekman. 1992.
\newblock An argument for basic emotions.
\newblock \emph{Cognition \& emotion}, 6(3-4):169--200.

\bibitem[{Ekman(2004)}]{ekman2004emotions}
Paul Ekman. 2004.
\newblock Emotions revealed.
\newblock \emph{Bmj}, 328(Suppl S5).

\bibitem[{Elyoseph et~al.(2023)Elyoseph, Hadar-Shoval, Asraf, and Lvovsky}]{elyoseph2023chatgpt}
Zohar Elyoseph, Dorit Hadar-Shoval, Kfir Asraf, and Maya Lvovsky. 2023.
\newblock Chatgpt outperforms humans in emotional awareness evaluations.
\newblock \emph{Frontiers in Psychology}, 14:1199058.

\bibitem[{Faul et~al.(2009)Faul, Erdfelder, Buchner, and Lang}]{faul2009statistical}
Franz Faul, Edgar Erdfelder, Axel Buchner, and Albert-Georg Lang. 2009.
\newblock Statistical power analyses using g* power 3.1: Tests for correlation and regression analyses.
\newblock \emph{Behavior research methods}, 41(4):1149--1160.

\bibitem[{Fu et~al.(2023)Fu, Inoue, Chu, and Kawahara}]{fu2023reasoning}
Yahui Fu, Koji Inoue, Chenhui Chu, and Tatsuya Kawahara. 2023.
\newblock Reasoning before responding: Integrating commonsense-based causality explanation for empathetic response generation.
\newblock \emph{arXiv preprint arXiv:2308.00085}.

\bibitem[{Goetz et~al.(2003)Goetz, Kiesler, and Powers}]{goetz2003matching}
Jennifer Goetz, Sara Kiesler, and Aaron Powers. 2003.
\newblock Matching robot appearance and behavior to tasks to improve human-robot cooperation.
\newblock In \emph{The 12th IEEE International Workshop on Robot and Human Interactive Communication, 2003. Proceedings. ROMAN 2003.}, pages 55--60. Ieee.

\bibitem[{Ioannidou and Konstantikaki(2008)}]{ioannidou2008empathy}
Flora Ioannidou and Vaya Konstantikaki. 2008.
\newblock Empathy and emotional intelligence: What is it really about?
\newblock \emph{International Journal of caring sciences}, 1(3):118.

\bibitem[{Kang(2021)}]{kang2021sample}
Hyun Kang. 2021.
\newblock Sample size determination and power analysis using the g* power software.
\newblock \emph{Journal of educational evaluation for health professions}, 18.

\bibitem[{Koufouli and Tollenaar(2016)}]{koufouli2016empathy}
Alexandra Koufouli and Marieke~S Tollenaar. 2016.
\newblock Empathy and emotional awareness: An interdisciplinary perspective.
\newblock \emph{Offenders no more: An interdisciplinary restorative justice dialogue}.

\bibitem[{Lane et~al.(1990)Lane, Quinlan, Schwartz, Walker, and Zeitlin}]{lane1990levels}
Richard~D Lane, Donald~M Quinlan, Gary~E Schwartz, Pamela~A Walker, and Sharon~B Zeitlin. 1990.
\newblock The levels of emotional awareness scale: A cognitive-developmental measure of emotion.
\newblock \emph{Journal of personality assessment}, 55(1-2):124--134.

\bibitem[{Laskar et~al.(2023)Laskar, Bari, Rahman, Bhuiyan, Joty, and Huang}]{laskarsystematic}
Md~Tahmid~Rahman Laskar, M~Saiful Bari, Mizanur Rahman, Md~Amran~Hossen Bhuiyan, Shafiq Joty, and Jimmy Huang. 2023.
\newblock A systematic study and comprehensive evaluation of {C}hat{GPT} on benchmark datasets.
\newblock In \emph{Findings of the Association for Computational Linguistics: ACL 2023}, pages 431--469, Toronto, Canada. Association for Computational Linguistics.

\bibitem[{Lee et~al.(2022)Lee, Lim, and Choi}]{lee2022does}
Young-Jun Lee, Chae-Gyun Lim, and Ho-Jin Choi. 2022.
\newblock Does gpt-3 generate empathetic dialogues? a novel in-context example selection method and automatic evaluation metric for empathetic dialogue generation.
\newblock In \emph{Proceedings of the 29th International Conference on Computational Linguistics}, pages 669--683.

\bibitem[{Liu et~al.(2023)Liu, McCoy, Wright, Carew, Genkins, Huang, Peterson, Steitz, and Wright}]{liu2023leveraging}
Siru Liu, Allison~B McCoy, Aileen~P Wright, Babatunde Carew, Julian~Z Genkins, Sean~S Huang, Josh~F Peterson, Bryan Steitz, and Adam Wright. 2023.
\newblock Leveraging large language models for generating responses to patient messages.
\newblock \emph{medRxiv}, pages 2023--07.

\bibitem[{Liu-Thompkins et~al.(2022)Liu-Thompkins, Okazaki, and Li}]{liu2022artificial}
Yuping Liu-Thompkins, Shintaro Okazaki, and Hairong Li. 2022.
\newblock Artificial empathy in marketing interactions: Bridging the human-ai gap in affective and social customer experience.
\newblock \emph{Journal of the Academy of Marketing Science}, 50(6):1198--1218.

\bibitem[{Montoya(2023)}]{montoya2023selecting}
Amanda~K Montoya. 2023.
\newblock Selecting a within-or between-subject design for mediation: Validity, causality, and statistical power.
\newblock \emph{Multivariate Behavioral Research}, 58(3):616--636.

\bibitem[{Moyers et~al.(2014)Moyers, Manuel, Ernst, Moyers, Manuel, Ernst, and Fortini}]{miti_4_2_1}
TB~Moyers, JK~Manuel, D~Ernst, T~Moyers, J~Manuel, D~Ernst, and C~Fortini. 2014.
\newblock Motivational interviewing treatment integrity coding manual 4.1 (miti 4.1).
\newblock \emph{Unpublished manual}.

\bibitem[{Peer et~al.(2022)Peer, Rothschild, Gordon, Evernden, and Damer}]{peer2022data}
Eyal Peer, David Rothschild, Andrew Gordon, Zak Evernden, and Ekaterina Damer. 2022.
\newblock Data quality of platforms and panels for online behavioral research.
\newblock \emph{Behavior Research Methods}, page~1.

\bibitem[{Plutchik(1984)}]{plutchik}
Robert Plutchik. 1984.
\newblock Emotions: A general psychoevolutionary theory.
\newblock \emph{Approaches to emotion}, 1984(197-219):2--4.

\bibitem[{Powell and Roberts(2017)}]{powell2017situational}
Philip~A Powell and Jennifer Roberts. 2017.
\newblock Situational determinants of cognitive, affective, and compassionate empathy in naturalistic digital interactions.
\newblock \emph{Computers in Human Behavior}, 68:137--148.

\bibitem[{Qian et~al.(2023)Qian, Zhang, and Liu}]{qian2023harnessing}
Yushan Qian, Wei-Nan Zhang, and Ting Liu. 2023.
\newblock Harnessing the power of large language models for empathetic response generation: Empirical investigations and improvements.
\newblock \emph{arXiv preprint arXiv:2310.05140}.

\bibitem[{Rashkin et~al.(2019)Rashkin, Smith, Li, and Boureau}]{empatheticdialogues}
Hannah Rashkin, Eric~Michael Smith, Margaret Li, and Y-Lan Boureau. 2019.
\newblock Towards empathetic open-domain conversation models: A new benchmark and dataset.
\newblock In \emph{Proceedings of the 57th Annual Meeting of the Association for Computational Linguistics}, pages 5370--5381, Florence, Italy. Association for Computational Linguistics.

\bibitem[{Rieffe and Camodeca(2016)}]{rieffe2016empathy}
Carolien Rieffe and Marina Camodeca. 2016.
\newblock Empathy in adolescence: Relations with emotion awareness and social roles.
\newblock \emph{British journal of developmental psychology}, 34(3):340--353.

\bibitem[{Schaaff et~al.(2023)Schaaff, Reinig, and Schlippe}]{schaaff2023exploring}
Kristina Schaaff, Caroline Reinig, and Tim Schlippe. 2023.
\newblock Exploring chatgpt's empathic abilities.
\newblock \emph{arXiv preprint arXiv:2308.03527}.

\bibitem[{Schuler et~al.(2016)Schuler, Mohnke, and Walter}]{schuler2016neurological}
Miriam Schuler, Sebastian Mohnke, and Henrik Walter. 2016.
\newblock The neurological basis of empathy and mimicry.
\newblock \emph{Emotional mimicry in social context}, pages 192--221.

\bibitem[{Semenick(1990)}]{ttest}
Doug Semenick. 1990.
\newblock Tests and measurements: The t-test.
\newblock \emph{Strength \& Conditioning Journal}, 12(1):36--37.

\bibitem[{Shaughnessy et~al.(2000)Shaughnessy, Zechmeister, and Zechmeister}]{shaughnessy2000research}
John~J Shaughnessy, Eugene~B Zechmeister, and Jeanne~S Zechmeister. 2000.
\newblock \emph{Research methods in psychology}.
\newblock McGraw-Hill.

\bibitem[{Skerry and Saxe(2015)}]{skerry}
Amy~E Skerry and Rebecca Saxe. 2015.
\newblock Neural representations of emotion are organized around abstract event features.
\newblock \emph{Current biology}, 25(15):1945--1954.

\bibitem[{Spreng et~al.(2009)Spreng, McKinnon, Mar, and Levine}]{spreng2009toronto}
R~Nathan Spreng, Margaret~C McKinnon, Raymond~A Mar, and Brian Levine. 2009.
\newblock The toronto empathy questionnaire: Scale development and initial validation of a factor-analytic solution to multiple empathy measures.
\newblock \emph{Journal of personality assessment}, 91(1):62--71.

\bibitem[{St et~al.(1989)St, Wold et~al.}]{anova}
Lars St, Svante Wold, et~al. 1989.
\newblock Analysis of variance (anova).
\newblock \emph{Chemometrics and intelligent laboratory systems}, 6(4):259--272.

\bibitem[{Stroessner and Benitez(2019)}]{stroessner2019social}
Steven~J Stroessner and Jonathan Benitez. 2019.
\newblock The social perception of humanoid and non-humanoid robots: Effects of gendered and machinelike features.
\newblock \emph{International Journal of Social Robotics}, 11:305--315.

\bibitem[{Svikhnushina and Pu(2022)}]{svikhnushina2022peace}
Ekaterina Svikhnushina and Pearl Pu. 2022.
\newblock Peace: A model of key social and emotional qualities of conversational chatbots.
\newblock \emph{ACM Transactions on Interactive Intelligent Systems}, 12(4):1--29.

\bibitem[{Vanderlyn et~al.(2021)Vanderlyn, Weber, Neumann, V{\"a}th, Meyer, and Vu}]{vanderlyn2021seemed}
Lindsey Vanderlyn, Gianna Weber, Michael Neumann, Dirk V{\"a}th, Sarina Meyer, and Ngoc~Thang Vu. 2021.
\newblock “it seemed like an annoying woman”: On the perception and ethical considerations of affective language in text-based conversational agents.
\newblock In \emph{Proceedings of the 25th Conference on Computational Natural Language Learning}, pages 44--57.

\bibitem[{Welivita and Pu(2020)}]{welivitapu2020taxonomy}
Anuradha Welivita and Pearl Pu. 2020.
\newblock A taxonomy of empathetic response intents in human social conversations.
\newblock In \emph{Proceedings of the 28th International Conference on Computational Linguistics}, pages 4886--4899, Barcelona, Spain (Online). International Committee on Computational Linguistics.

\bibitem[{Welivita and Pu(2023)}]{boosting}
Anuradha Welivita and Pearl Pu. 2023.
\newblock Boosting distress support dialogue responses with motivational interviewing strategy.
\newblock In \emph{Findings of the Association for Computational Linguistics: ACL 2023}, pages 5411--5432, Toronto, Canada. Association for Computational Linguistics.

\bibitem[{Zhao et~al.(2023)Zhao, Zhao, Lu, Wang, Tong, and Qin}]{zhao2023chatgpt}
Weixiang Zhao, Yanyan Zhao, Xin Lu, Shilong Wang, Yanpeng Tong, and Bing Qin. 2023.
\newblock Is chatgpt equipped with emotional dialogue capabilities?
\newblock \emph{arXiv preprint arXiv:2304.09582}.

\bibitem[{Ziyu et~al.(2023)Ziyu, Qiguang, Longxuan, Mingda, Yi, Yushan, Haopeng, Weinan, and Liu}]{ziyulens}
Zhuang Ziyu, Chen Qiguang, Ma~Longxuan, Li~Mingda, Han Yi, Qian Yushan, Bai Haopeng, Zhang Weinan, and Ting Liu. 2023.
\newblock Through the lens of core competency: Survey on evaluation of large language models.
\newblock In \emph{Proceedings of the 22nd Chinese National Conference on Computational Linguistics (Volume 2: Frontier Forum)}, pages 88--109, Harbin, China.

\end{thebibliography}

\appendix

\section{Distribution of Emotions}
\label{app:emotions}

\begin{figure*}
\centering
\includegraphics[width=0.85\textwidth]{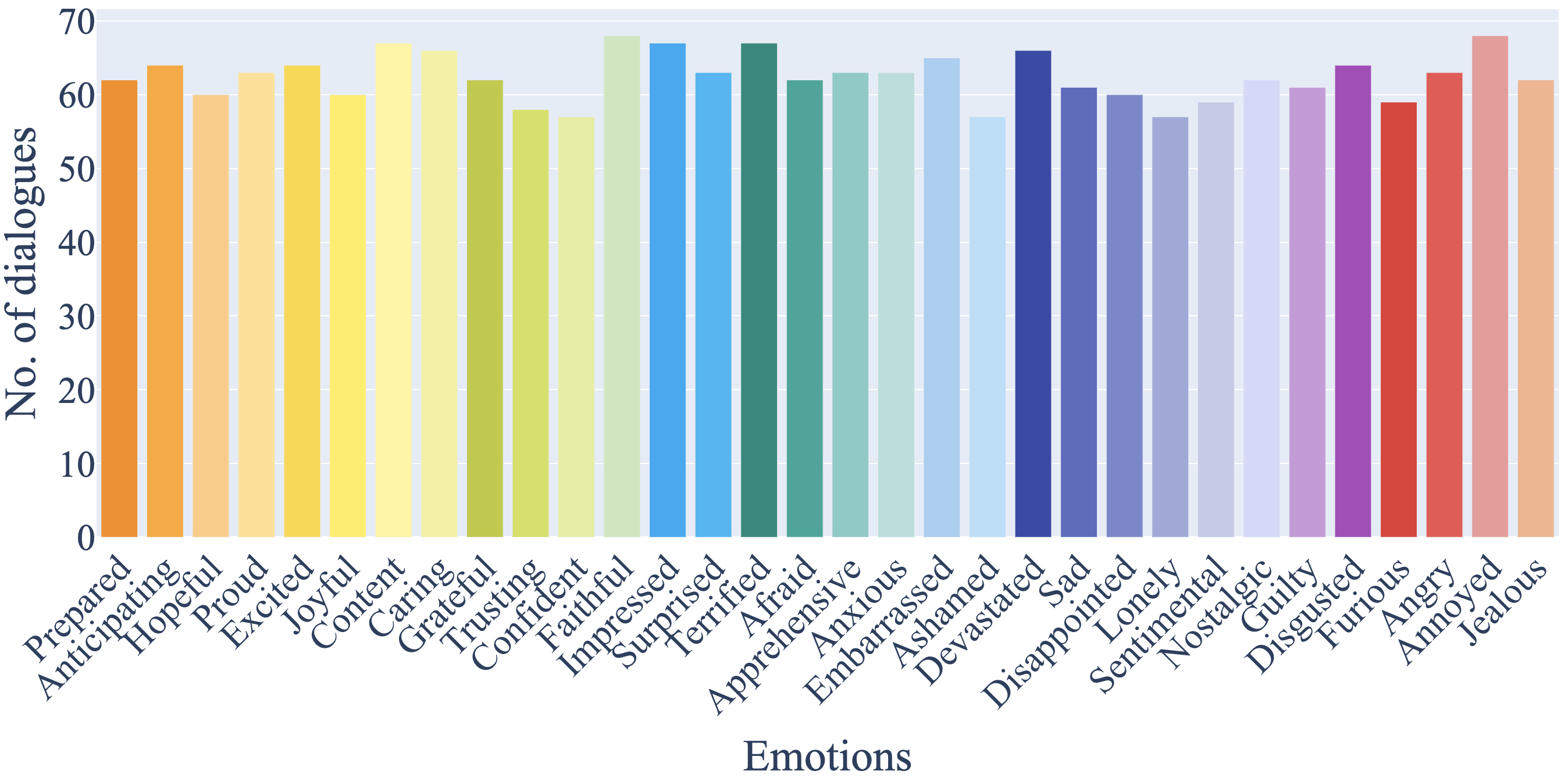}
\caption{Distribution of the dialogue prompt-response pairs sampled from the EmpatheticDialogues dataset across the 32 positive and negative emotions.}
\label{fig:emotions}
\end{figure*}

Figure \ref{fig:emotions} shows the distribution of the dialogue prompt-response pairs sampled from the EmpatheticDialogues dataset across the 32 positive and negative emotions. Table \ref{tab:emotions} shows the counts and the percentages of dialogue prompt-response pairs in the dataset corresponding to each emotion. It can be noted that the prompt-response pairs are more or less equally distributed across the 32 emotions. 

\section{Experiment Parameters}
\label{app:chatgpt}

The following are the key parameters used when querying OpenAI's GPT-4 to generate responses to the emotional dialogue prompts: \texttt{model=gpt-4}; \texttt{temperature=0}; \texttt{top\_p=1}; \texttt{frequency\_penalty=0}; and \texttt{presence\_penalty=0}. All the experiments were conducted on a MacBook Pro machine having a 2.3 GHz Quad-Core Intel Core i5 processor and 8 GB memory.

\begin{table}[ht!]
\small
\centering
\begin{tabular}{l c c}
\toprule

\textbf{Emotion} & \textbf{\# dialogues} & \textbf{\% of dialogues} \\

\midrule

\textbf{\underline{Positive emotions:}} & \textbf{881} & \textbf{44.05\%} \vspace{1mm} \\

Prepared & 62 & 3.10\% \\
Anticipating & 64 & 3.20\% \\
Hopeful & 60 & 3.00\% \\
Proud & 63 & 3.15\% \\
Excited & 64 & 3.20\% \\
Joyful & 60 & 3.00\% \\
Content & 67 & 3.35\% \\
Caring & 66 & 3.30\% \\
Grateful & 62 & 3.10\% \\
Trusting & 58 & 2.90\% \\
Confident & 57 & 2.85\% \\
Faithful & 68 & 3.40\% \\
Impressed & 67 & 3.35\% \\
Surprised & 63 & 3.15\% \vspace{1mm} \\

\textbf{\underline{Negative emotions:}} & \textbf{1119} & \textbf{55.95\%} \vspace{1mm} \\

Terrified & 67 & 3.35\% \\
Afraid & 62 & 3.10\% \\
Apprehensive & 63 & 3.15\% \\
Anxious & 63 & 3.15\% \\
Embarrassed & 65 & 3.25\% \\
Ashamed & 57 & 2.85\% \\
Devastated & 66 & 3.30\% \\
Sad & 61 & 3.05\% \\
Disappointed & 60 & 3.00\% \\
Lonely & 57 & 2.85\% \\
Sentimental & 59 & 2.95\% \\
Nostalgic & 62 & 3.10\% \\
Guilty & 61 & 3.05\% \\
Disgusted & 64 & 3.20\% \\
Furious & 59 & 2.95\% \\
Angry & 63 & 3.15\% \\
Annoyed & 68 & 3.40\% \\
Jealous & 62 & 3.10\% \\

\bottomrule
\end{tabular}
\caption{The counts and percentages of dialogue prompt-response pairs in the dataset corresponding to each emotion.}
\label{tab:emotions}
\end{table}

\section{Toronto Empathy Questionnaire}
\label{app:teq}

Table \ref{tab:teq} shows the questions in the Toronto Empathy Questionnaire (TEQ) \cite{spreng2009toronto} that were asked from the participants. Responses to the questions were scored according to the following scale for positively worded questions: Never = 0; Rarely = 1; Sometimes = 2; Often = 3; Always = 4. The negatively worded questions indicated are reverse-scored. Scores were summed to derive the evaluator's propensity to empathize.

\begin{table*}[ht!]
\small
\centering
\begin{tabular}{r l}
\toprule

1. & \textit{When someone else is feeling excited, I tend to get excited too} \\
2. & \textit{Other people’s misfortunes do not disturb me a great deal\textsuperscript{*}} \\
3. & \textit{It upsets me to see someone being treated disrespectfully} \\
4. & \textit{I remain unaffected when someone close to me is happy\textsuperscript{*}} \\
5. & \textit{I enjoy making other people feel better} \\
6. & \textit{I have tender, concerned feelings for people less fortunate than me} \\
7. & \textit{When a friend starts to talk about his or her problems, I try to steer the conversation towards something else\textsuperscript{*}} \\
8. & \textit{I can tell when others are sad even when they do not say anything} \\
9. & \textit{I find that I am “in tune” with other people’s moods} \\
10. & \textit{I do not feel sympathy for people who cause their own serious illnesses\textsuperscript{*}} \\
11. & \textit{I become irritated when someone cries\textsuperscript{*}} \\
12. & \textit{I am not really interested in how other people feel\textsuperscript{*}} \\
13. & \textit{I get a strong urge to help when I see someone who is upset} \\
14. & \textit{When I see someone being treated unfairly, I do not feel very much pity for them\textsuperscript{*}} \\
15. & \textit{I find it silly for people to cry out of happiness\textsuperscript{*}} \\
16. & \textit{When I see someone being taken advantage of, I feel kind of protective towards him or her} \\

\bottomrule
\end{tabular}
\caption{The Toronto Empathy Questionnaire \cite{spreng2009toronto}. \textsuperscript{*}Negatively worded reverse scale questions.}
\label{tab:teq}
\end{table*}

\section{Task Interfaces}
\label{app:task_interfaces}

\begin{figure}
  \fbox{\includegraphics[width=\linewidth]{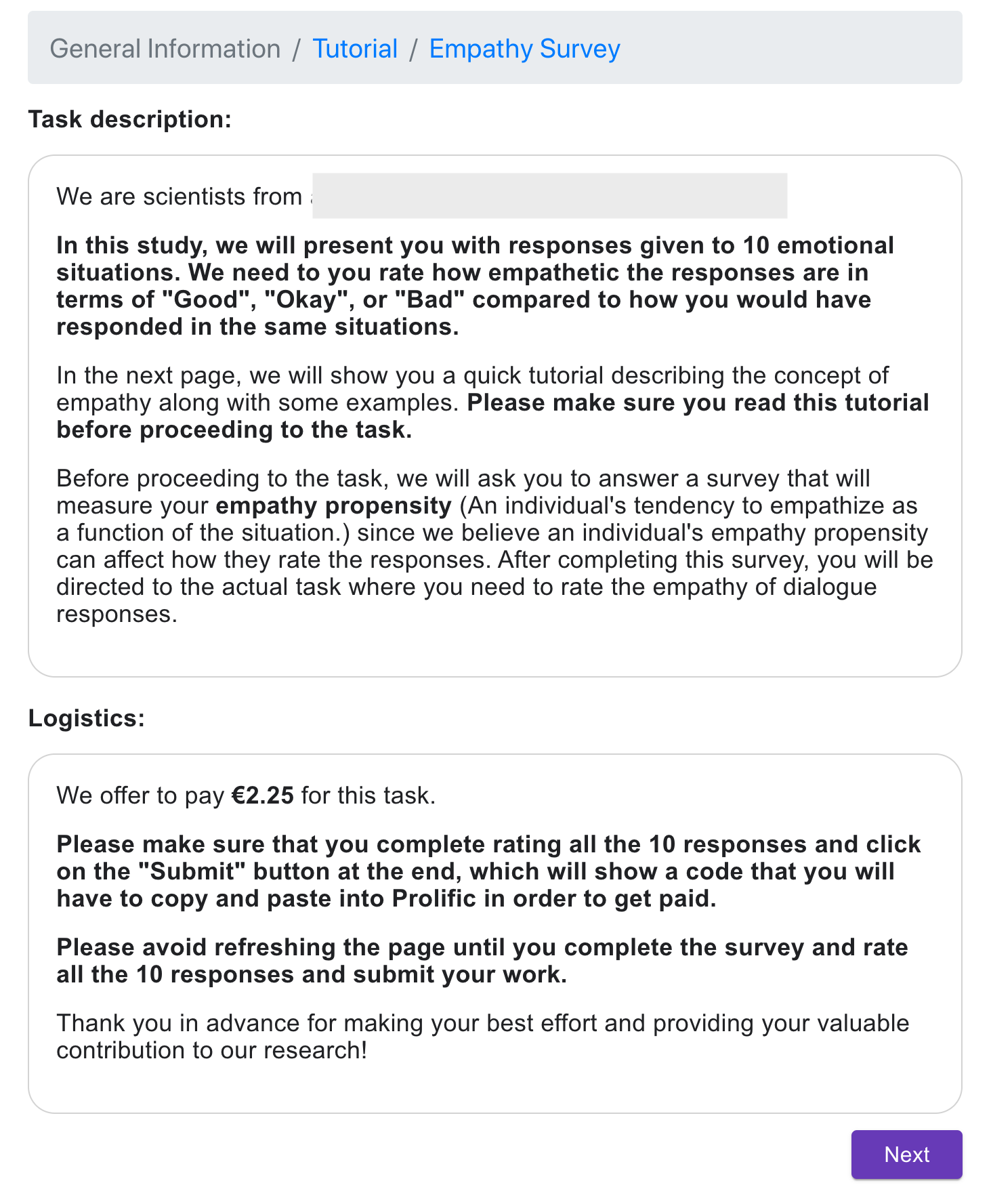}}
  \caption{The description of the task.}
  \label{fig:task_info}
\end{figure}

\begin{figure}
  \fbox{\includegraphics[width=\linewidth]{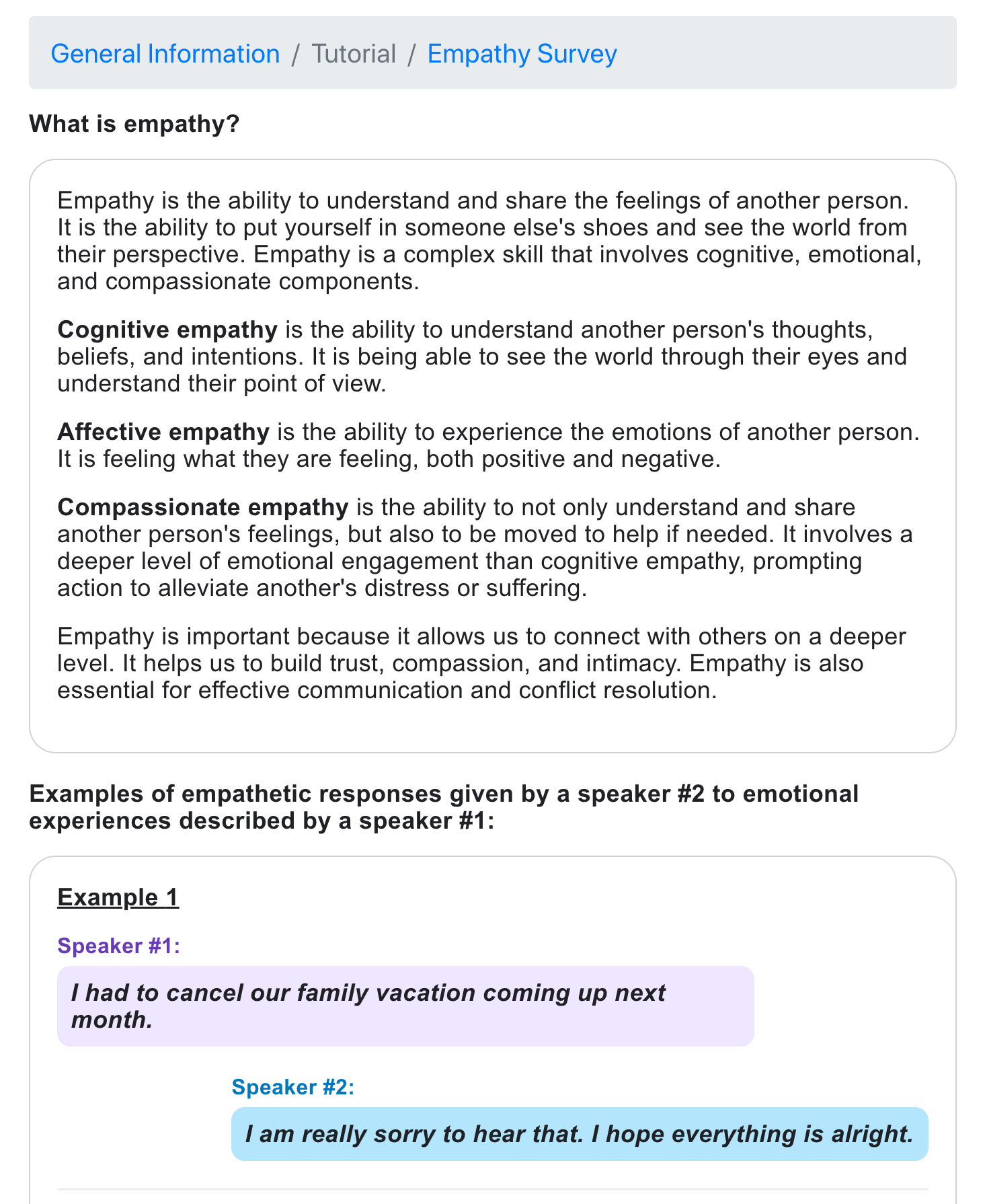}}
  \caption{The tutorial.}
  \label{fig:tutorial}
\end{figure}

\begin{figure}
  \fbox{\includegraphics[width=\linewidth]{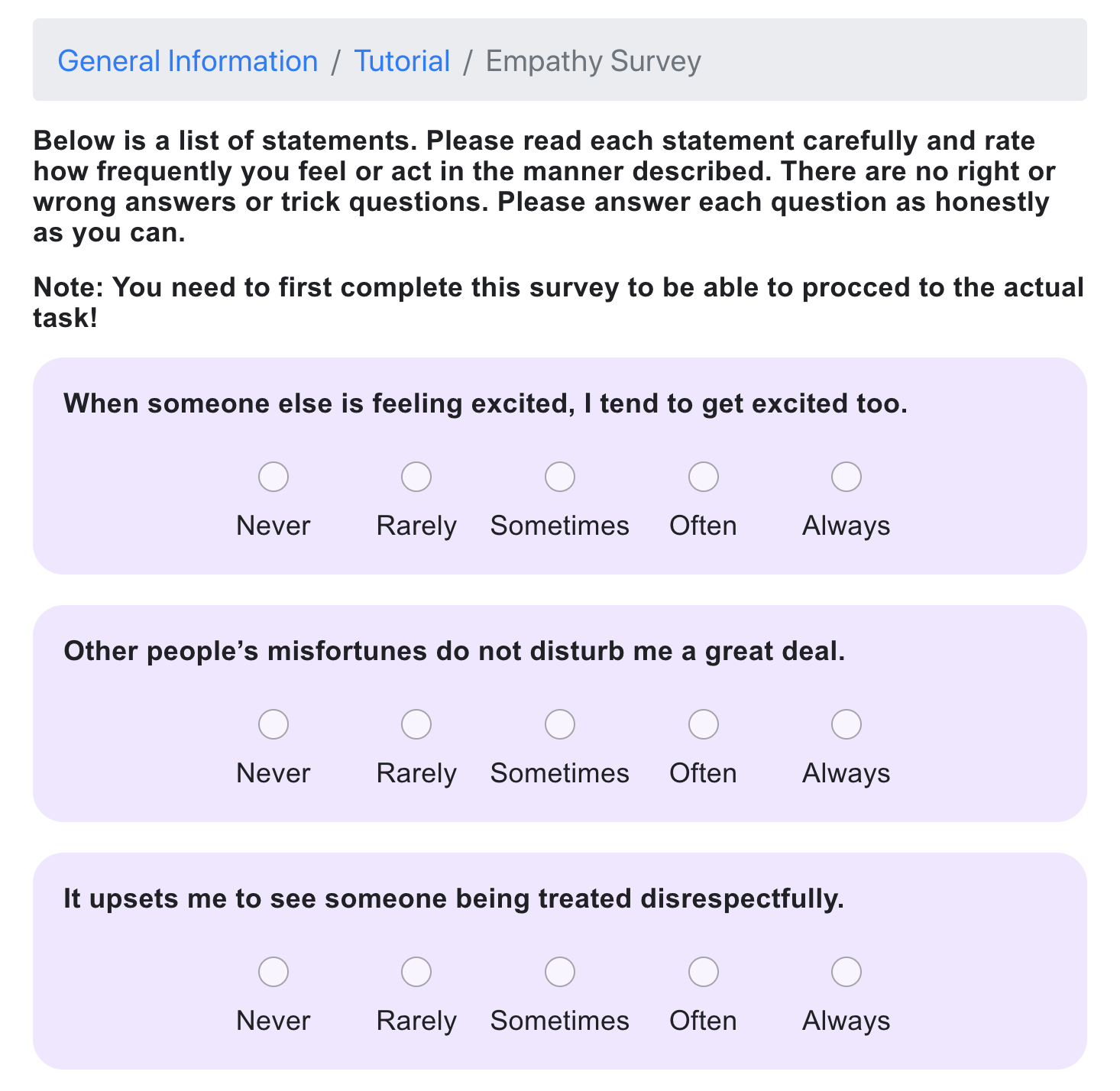}}
  \caption{The Toronto Empathy Questionnaire.}
  \label{fig:teq}
\end{figure}

\begin{figure}
  \fbox{\includegraphics[width=\linewidth]{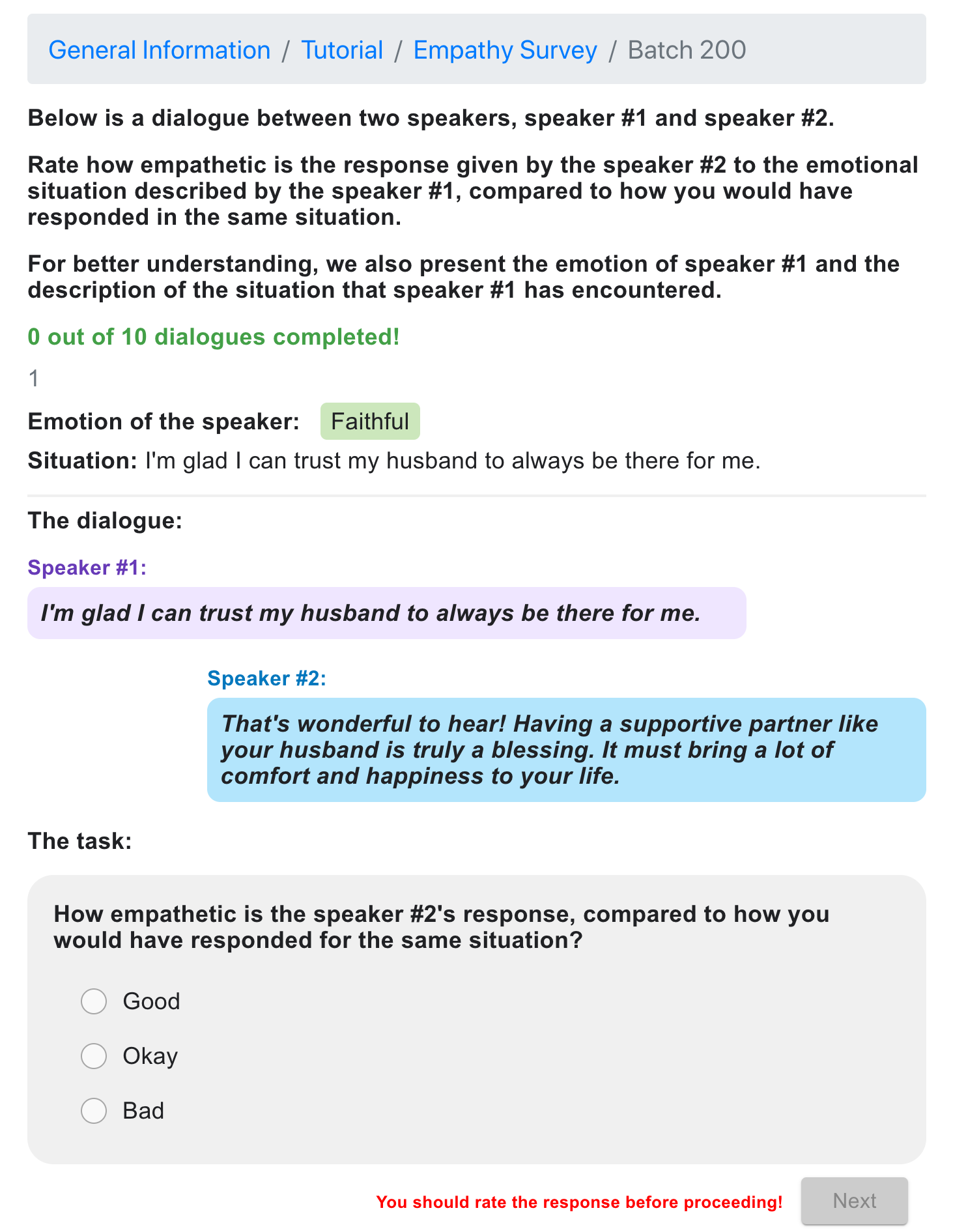}}
  \caption{The task interface for rating responses in terms of empathy.}
  \label{fig:task}
\end{figure}

Figures \ref{fig:task_info}, \ref{fig:tutorial}, \ref{fig:teq} and \ref{fig:task} show the task interfaces corresponding to the description of the task, the tutorial presented to the crowd workers, the Toronto Empathy Questionnaire, and the response rating task, respectively. 

\section{Determing the Effect Size}
\label{app:effect_size}

Jacob Cohen, a notable psychologist and statistician, established benchmarks for interpreting effect sizes in statistical tests, including analysis of variance (ANOVA) and chi-square tests \cite{cohen1992quantitative}. These benchmarks serve as a general framework for evaluating the practical significance of the effects observed in these tests. For ANOVA, Cohen's F is a commonly used measure of effect size, representing the standardized difference between means. Cohen suggested the following benchmarks for interpreting the magnitude of Cohen's F: small effect: 0.10; medium effect: 0.25; and large effect: 0.40. 

We selected medium effect sizes to compute the required minimum sample size because a medium effect size can sensitively detect differences in empathy levels between the responses of humans and ChatGPT, which are expected to be noticeable but not extremely large. Also, a study conducted with a medium effect size can detect differences that are subtle yet practically meaningful, without requiring an excessively large sample size.

\section{One-way ANOVA Test Results}
\label{app:anova}

\begin{figure*}
  \centering
  \includegraphics[width=\textwidth]{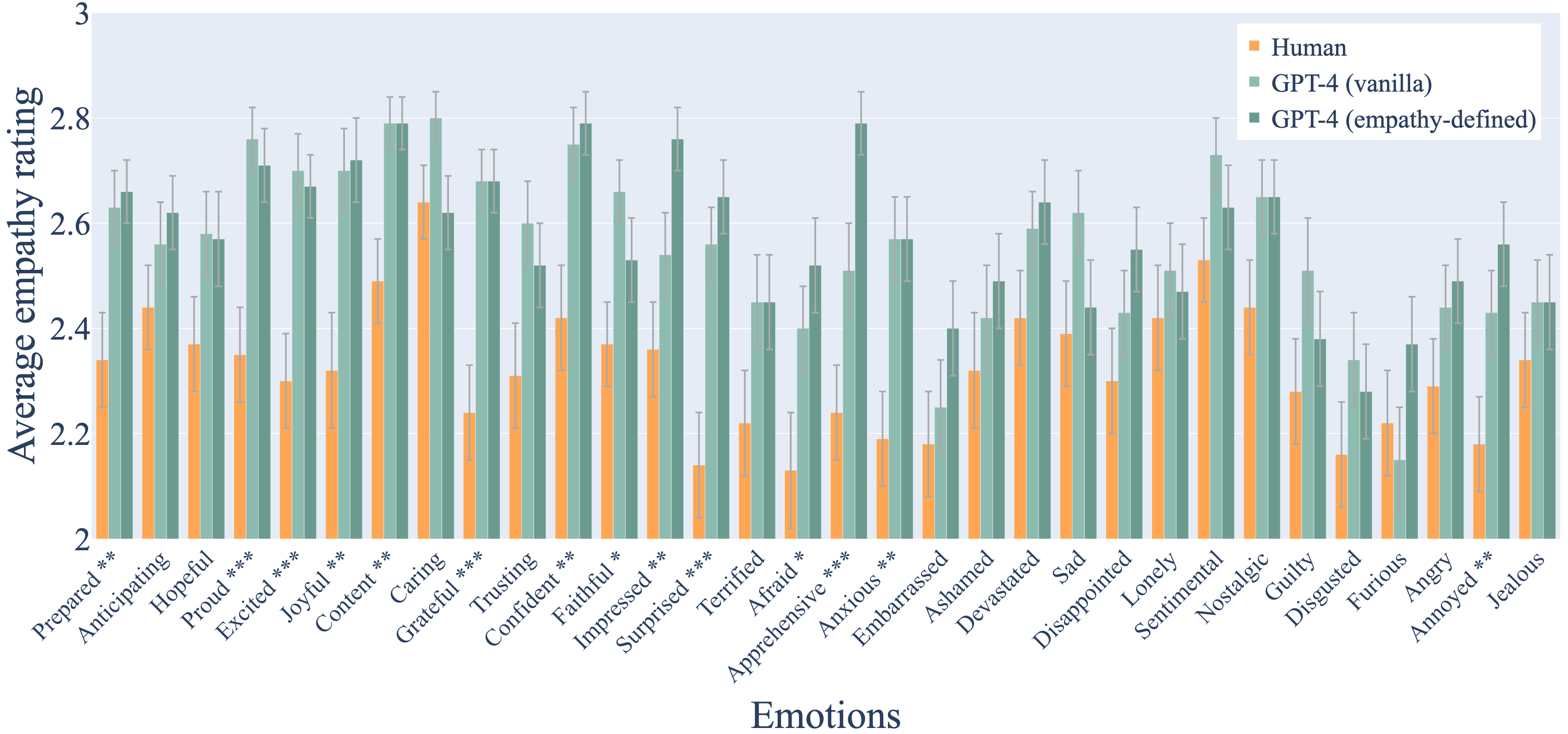}
  \caption{Average empathy ratings corresponding to the human’s and the two versions of GPT-4’s responses across the 32 positive and negative emotions. Error bars are calculated using the standard errors for each. We have indicated ***, **, and * in front of the emotion in the x-axis to indicate results that are statistically significant (***, **, and * indicate p < 0.001, p < 0.01, and p < 0.5, respectively).}
  \label{fig:main_results_emotions}
\end{figure*}

The statistical one-way ANOVA test results corresponding to the average empathy ratings of the responses generated by the humans and the two variants of GPT-4 are denoted in Table \ref{table:anova}. Table \ref{table:anova_emotions} denotes the statistical one-way ANOVA test results corresponding to each of the 32 emotions the dialogue prompts used for testing are based on. Figure \ref{fig:main_results_emotions} visualizes the average empathy ratings corresponding to the human’s and the two variants of GPT-4’s responses across the 32 positive and negative emotions. All values in Table \ref{table:anova} are statistically significant as denoted by the low p-values (p < 0.001). Only some values in Table \ref{table:anova_emotions} are statistically significant mainly due to the insufficient sample size corresponding to each emotion. Nevertheless, it can be observed that the responses generated by GPT-4 (empathy-defined) score the highest average empathy rating out of the three response groups across most emotions. 

\begin{table*}[ht!]
\small
\centering
\begin{tabular}{l | r r r | r l}
\toprule

\textbf{Emotion} & \multicolumn{1}{c}{\textbf{Human}} & \multicolumn{1}{c}{\textbf{GPT-4}} & \multicolumn{1}{c}{\textbf{GPT-4}} & \multicolumn{1}{c}{\textbf{F-value}} & \multicolumn{1}{c}{\textbf{p-value}} \\

\textbf{} & \textbf{} & \multicolumn{1}{c}{\textbf{(vanilla)}} & \multicolumn{1}{c}{\textbf{(empathy-}} & \textbf{} & \textbf{} \\

\textbf{} & \textbf{} & \textbf{} & \multicolumn{1}{c}{\textbf{defined)}} & \textbf{} & \textbf{} \\

\midrule

All emotions & 2.32 ± 0.02 & 2.56 ± 0.01 & \textbf{2.58 ± 0.01} & \textbf{88.849} & \textbf{9.418e-39***} \\
Positive emotions & 2.36 ± 0.02 & \textbf{2.67 ± 0.02} & 2.66 ± 0.02 & \textbf{69.558} & \textbf{3.632e-30***} \\
Negative emotions & 2.29 ± 0.02 & 2.47 ± 0.02 & \textbf{2.51 ± 0.02} & \textbf{30.971} & \textbf{4.698e-14***} \\

\bottomrule

\end{tabular}
\caption{Statistical one-way ANOVA test results corresponding to the average empathy ratings of the human's and GPT-4's responses (based on the two prompts). Generally, any F-value greater than +2 or less than -2 is acceptable. The higher the F-value, the greater the confidence we have in the coefficient as a predictor.}
\label{table:anova}
\end{table*}

\begin{table*}[ht!]
\small
\centering
\begin{tabular}{l | r r r | r l}
\toprule

\textbf{Emotion} & \multicolumn{1}{c}{\textbf{Human}} & \multicolumn{1}{c}{\textbf{GPT-4}} & \multicolumn{1}{c}{\textbf{GPT-4}} & \multicolumn{1}{c}{\textbf{F-value}} & \multicolumn{1}{c}{\textbf{p-value}} \\

\textbf{} & \textbf{} & \multicolumn{1}{c}{\textbf{(vanilla)}} & \multicolumn{1}{c}{\textbf{(empathy-}} & \textbf{} & \textbf{} \\

\textbf{} & \textbf{} & \textbf{} & \multicolumn{1}{c}{\textbf{defined)}} & \textbf{} & \textbf{} \\

\midrule

\multicolumn{6}{l}{\textbf{Positive emotions:}\vspace{1mm}}\\

Prepared  &  2.34 ± 0.09  &  2.63 ± 0.07  &  \textbf{2.66 ± 0.06}  &  \textbf{5.406}  &  \textbf{0.005**}\\
Anticipating  &  2.44 ± 0.08  &  2.56 ± 0.08  &  \textbf{2.62 ± 0.07}  &  1.480  &  0.230\\
Hopeful  &  2.37 ± 0.09  &  \textbf{2.58 ± 0.08}  &  2.57 ± 0.09  &  1.854  &  0.160\\
Proud  &  2.35 ± 0.09  &  \textbf{2.76 ± 0.06}  &  2.71 ± 0.07  &  \textbf{8.708}  &  \textbf{0.000***}\\
Excited  &  2.30 ± 0.09  &  \textbf{2.70 ± 0.07}  &  2.67 ± 0.06  &  \textbf{8.728}  &  \textbf{0.000***}\\
Joyful  &  2.32 ± 0.11  &  2.70 ± 0.08  &  \textbf{2.72 ± 0.08}  &  \textbf{6.650}  &  \textbf{0.002**}\\
Content  &  2.49 ± 0.08  &  \textbf{2.79 ± 0.05}  &  \textbf{2.79 ± 0.05}  &  \textbf{6.925}  &  \textbf{0.001**}\\
Caring  &  2.64 ± 0.07  &  \textbf{2.80 ± 0.05}  &  2.62 ± 0.07  &  2.460  &  0.088\\
Grateful  &  2.24 ± 0.09  &  \textbf{2.68 ± 0.06}  &  \textbf{2.68 ± 0.06}  &  \textbf{11.126}  &  \textbf{0.000***}\\
Trusting  &  2.31 ± 0.10  &  \textbf{2.60 ± 0.08}  &  2.52 ± 0.08  &  3.010  &  0.052\\
Confident  &  2.42 ± 0.1  &  2.75 ± 0.07  &  \textbf{2.79 ± 0.06}  &  \textbf{6.835}  &  \textbf{0.001**}\\
Faithful  &  2.37 ± 0.08  &  \textbf{2.66 ± 0.06}  &  2.53 ± 0.08  &  \textbf{3.904}  &  \textbf{0.022*}\\
Impressed  &  2.36 ± 0.09  &  2.54 ± 0.08  &  \textbf{2.76 ± 0.06}  &  \textbf{6.912}  &  \textbf{0.001**}\\
Surprised  &  2.14 ± 0.10  &  2.56 ± 0.07  &  \textbf{2.65 ± 0.07}  &  \textbf{11.010}  &  \textbf{0.000***}\vspace{1mm}\\

\multicolumn{6}{l}{\textbf{Negative emotions:}\vspace{1mm}}\\

Terrified  &  2.22 ± 0.10  &  \textbf{2.45 ± 0.09}  &  \textbf{2.45 ± 0.09}  &  2.001  &  0.138\\
Afraid  &  2.13 ± 0.11  &  2.4 ± 0.08  &  \textbf{2.52 ± 0.09}  &  \textbf{4.618}  &  \textbf{0.011*}\\
Apprehensive  &  2.24 ± 0.09  &  2.51 ± 0.09  &  \textbf{2.79 ± 0.06}  &  \textbf{12.306}  &  \textbf{0.000***}\\
Anxious  &  2.19 ± 0.09  &  \textbf{2.57 ± 0.08}  &  \textbf{2.57 ± 0.08}  &  \textbf{6.548}  &  \textbf{0.002**}\\
Embarrassed  &  2.18 ± 0.10  &  2.25 ± 0.09  &  \textbf{2.40 ± 0.09}  &  1.354  &  0.261\\
Ashamed  &  2.32 ± 0.11  &  2.42 ± 0.10  &  \textbf{2.49 ± 0.09}  &  0.790  &  0.455\\
Devastated  &  2.42 ± 0.09  &  2.59 ± 0.07  &  \textbf{2.64 ± 0.08}  &  1.881  &  0.155\\
Sad  &  2.39 ± 0.10  &  \textbf{2.62 ± 0.08}  &  2.44 ± 0.09  &  1.783  &  0.171\\
Disappointed  &  2.30 ± 0.10  &  2.43 ± 0.08  &  \textbf{2.55 ± 0.08}  &  1.974  &  0.142\\
Lonely  &  2.42 ± 0.10  &  \textbf{2.51 ± 0.09}  &  2.47 ± 0.09  &  0.221  &  0.802\\
Sentimental  &  2.53 ± 0.08  &  \textbf{2.73 ± 0.07}  &  2.63 ± 0.08  &  1.765  &  0.174\\
Nostalgic  &  2.44 ± 0.09  &  \textbf{2.65 ± 0.07}  &  \textbf{2.65 ± 0.07}  &  2.613  &  0.076\\
Guilty  &  2.28 ± 0.10  &  \textbf{2.51 ± 0.10}  &  2.38 ± 0.09  &  1.429  &  0.242\\
Disgusted  &  2.16 ± 0.10  &  \textbf{2.34 ± 0.09}  &  2.28 ± 0.09  &  1.083  &  0.341\\
Furious  &  2.22 ± 0.1  &  2.15 ± 0.10  &  \textbf{2.37 ± 0.09}  &  1.313  &  0.272\\
Angry  &  2.29 ± 0.09  &  2.44 ± 0.08  &  \textbf{2.49 ± 0.08}  &  1.707  &  0.184\\
Annoyed  &  2.18 ± 0.09  &  2.43 ± 0.08  &  \textbf{2.56 ± 0.08}  &  \textbf{5.297}  &  \textbf{0.006**}\\
Jealous  &  2.34 ±0.09  &  \textbf{2.45 ± 0.08}  &  \textbf{2.45 ± 0.09}  &  0.584  &  0.559\\
\bottomrule
\end{tabular}
\caption{Statistical one-way ANOVA test results corresponding to the average empathy ratings of the human's and GPT-4's responses (based on the two prompts) to each of the 32 emotions the dialogue prompts are based on.}
\label{table:anova_emotions}
\end{table*}

\begin{figure}[h!]
\centering
\subfloat[\centering All emotions]
{{\includegraphics[width=\linewidth]{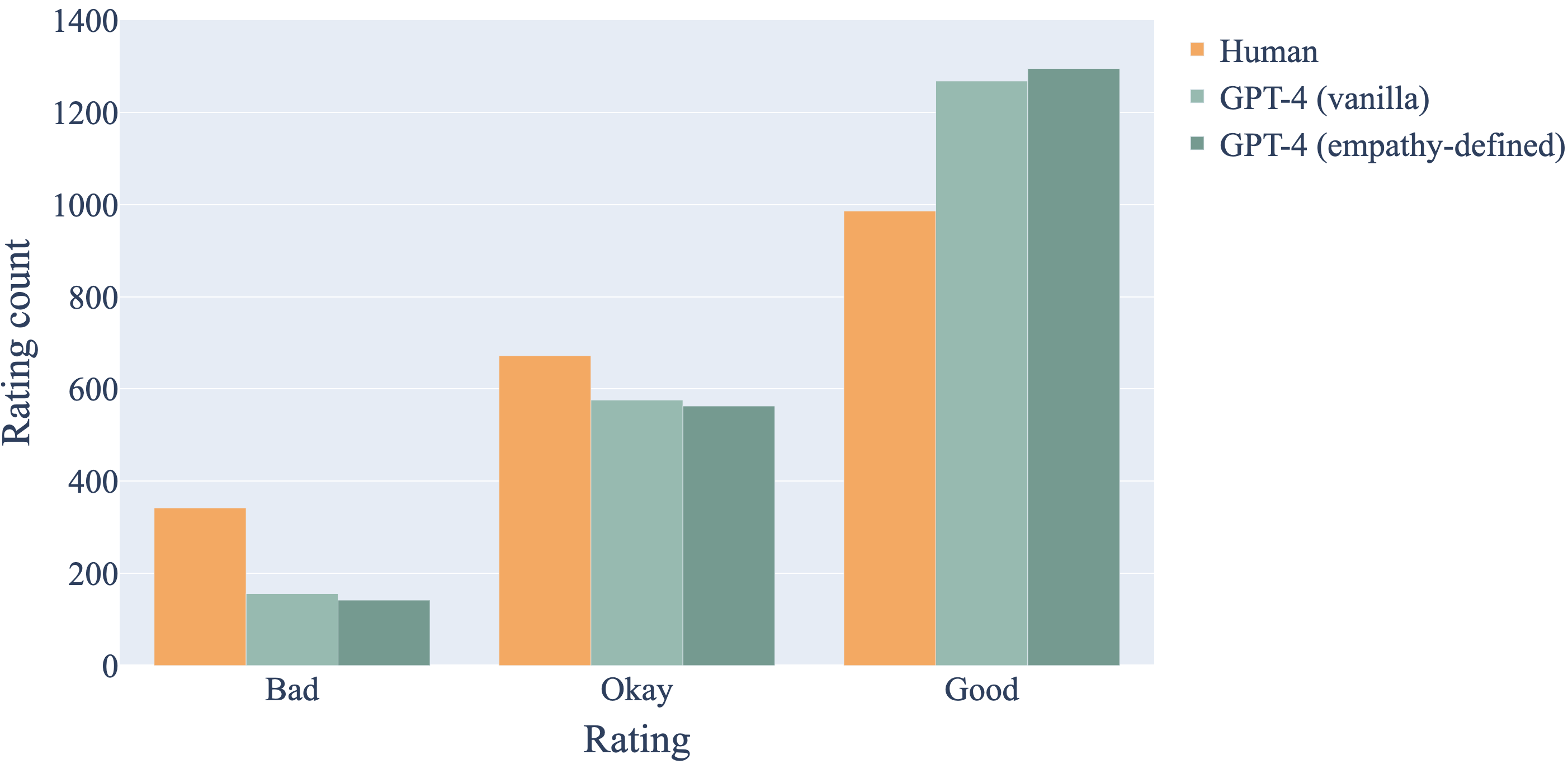}}}

\subfloat[\centering Positive emotions]
{{\includegraphics[width=\linewidth]{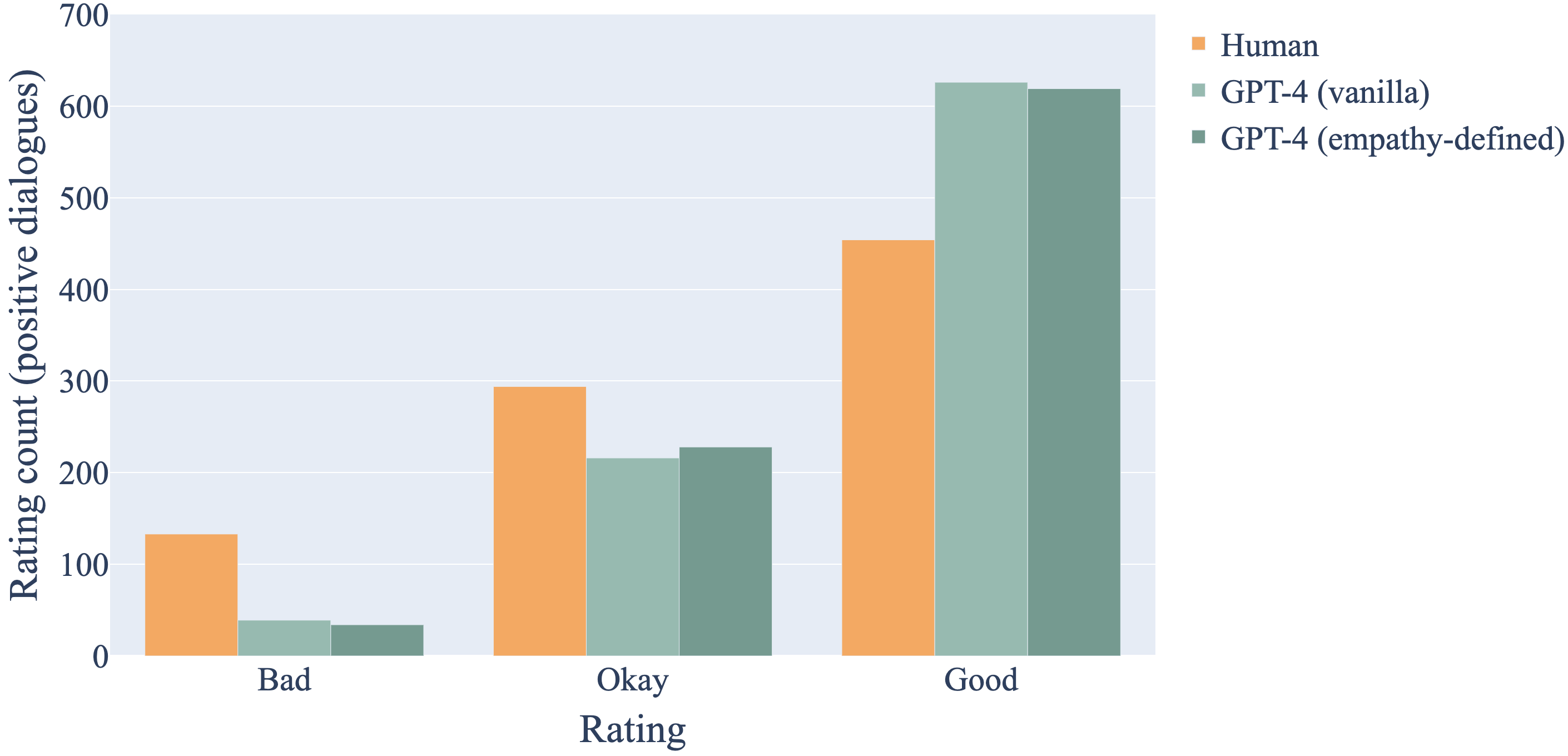}}}

\subfloat[\centering Negative emotions]
{{\includegraphics[width=\linewidth]{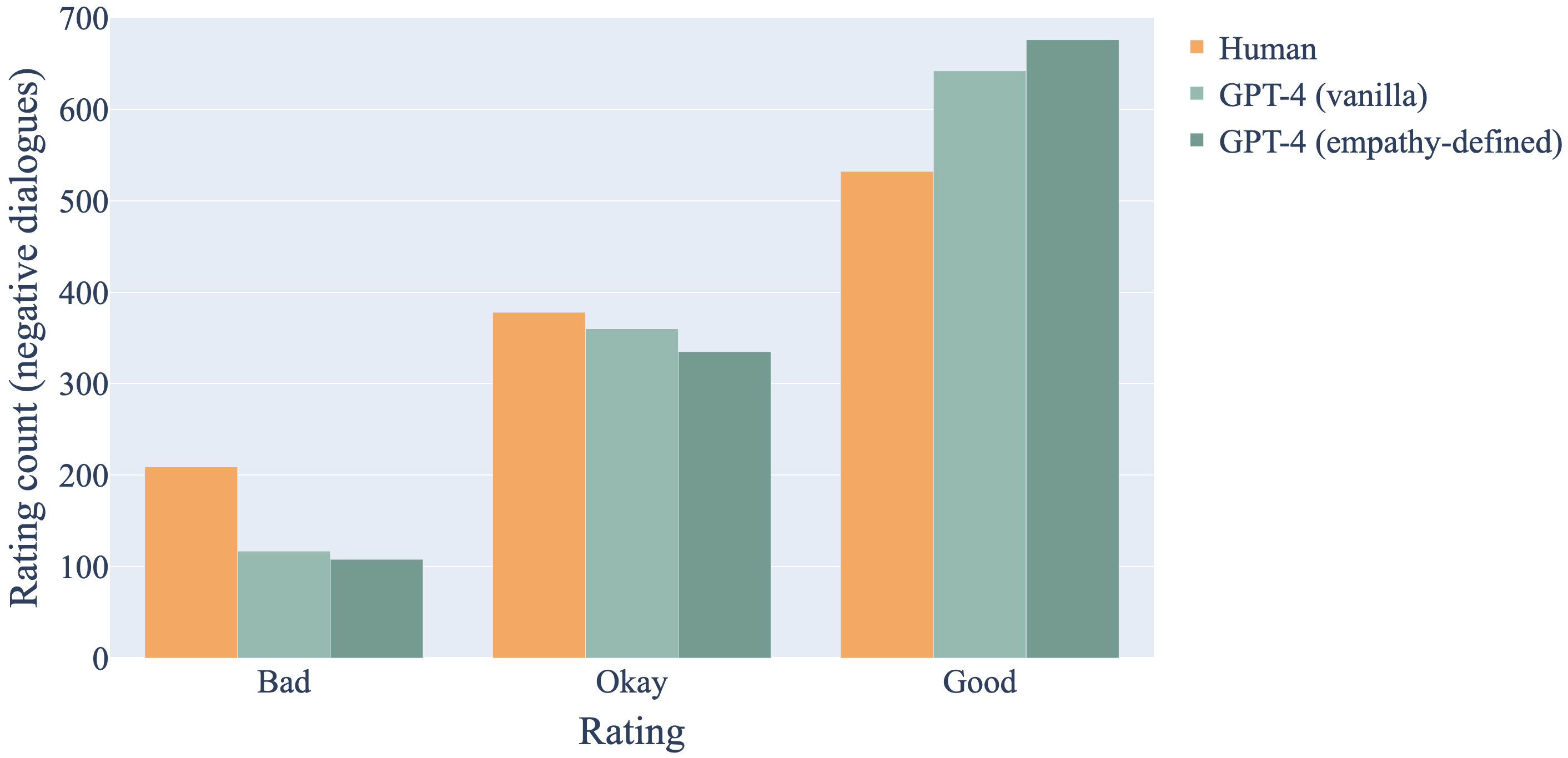}}}

\caption{The number of \textit{Bad}, \textit{Okay}, and \textit{Good} empathy ratings of the human’s and GPT-4’s responses (based on the two prompts) when responding to (a) all emotions (b) positive emotions, and (c) negative emotions.}
\label{fig:bars}
\end{figure}


\section{Chi-square Test of Independence Results}
\label{app:chi_square}

Another type of statistical test that can be used to analyze the results is \textbf{Chi-square test of independence} that tests whether there is any statistically significant difference between the proportion of \textit{Bad}, \textit{Okay}, and \textit{Good} ratings of the three groups. For the chi-square test of independence with a medium effect size (0.3), a significance level ($\alpha$) of 0.05, and a power (1-$\beta$) of 0.95, the minimal total sample size required is 207 (i.e. 69 participants per group). When statistically analyzing the differences in empathy ratings when responding to positive and negative emotions separately, the minimal sample size required becomes twice the sample size suggested above (i.e. 138 participants per group). Since we recruited 200 participants per group, which is sufficiently above the minimal sample size required for this test, the chi-square test of independence can also be applied to analyze the results obtained from this study. 

Table \ref{table:chi} shows the proportions of \textit{Bad}, \textit{Okay}, and \textit{Good} empathy ratings scored by the humans' and the two variants of GPT-4's responses. They are visualized in Figure \ref{fig:bars}. Similar to the case with average empathy ratings, the responses generated by GPT-4 that uses the empathy-defined prompt scores the highest number of \textit{Good} ratings when responding to all emotions as well as negative emotions, whereas the responses generated by GPT-4 that uses the generic prompt scores the highest number of \textit{Good} ratings when responding to positive emotions. The least number of \textit{Bad} responses are scored by GPT-4 based on the empathy-defined prompt when responding to all, positive, and negative emotions alike. The statistical ${\chi}^2$ and p-values obtained for the ${\chi}^2$ test comparing the proportions of \textit{Bad}, \textit{Okay}, and \textit{Good} empathy ratings of the three groups indicate that there is an extremely statistically significant difference (p < 0.001) between the proportions of the empathy ratings of GPT-4's responses compared to the humans'. However, similar to the case with average empathy ratings, the difference between the proportions of \textit{Bad}, \textit{Okay}, and \textit{Good} empathy ratings of GPT-4 based on the generic and empathy-defined prompts is not very statistically significant as indicated by the low ${\chi}^2$ (${\chi}^2$ < 5.991 --- critical value) and high p-values (p > 0.05) in Table \ref{table:chi_2}. 

\begin{table*}[h!]
\small
\centering
\begin{tabular}{l l | c c c | c c}
\toprule
 & \textbf{Type of response} & \textbf{Bad} & \textbf{Okay} & \textbf{Good} & \textbf{${\chi}^2$-value} & \textbf{p-value} \\
\midrule
All emotions & Human & 342 & 672 & 986 & \textbf{178.122} & \textbf{1.888e-37***} \\
& GPT-4 (vanilla) & 156 & 576 & 1268 & & \\
& GPT-4 (empathy-defined) &  \textbf{142} & 563 & \textbf{1295} & & \vspace{2mm}\\
Positive emotions & Human & 133 & 294 & 454 & \textbf{138.399} & \textbf{6.213e-29***} \\
& GPT-4 (vanilla) & 39 & 216 & \textbf{626} & & \\
& GPT-4 (empathy defined) & \textbf{34} &228 & 619 & & \vspace{2mm}\\
Negative emotions &  Human & 209 & 378 & 532 & \textbf{64.175} & \textbf{3.839e-13***} \\
& GPT-4 (vanilla) & 117 & 360 & 642 & & \\
& GPT-4 (empathy defined) & \textbf{108} & 335 & \textbf{676} & & \\
\bottomrule
\end{tabular}
\caption{Statistical Chi-square test results corresponding to the proportions of \textit{Bad}, \textit{Okay}, and \textit{Good} empathy ratings of the human's and GPT-4's responses (based on the two prompts). The critical value of the ${\chi}^2$ distribution in this case is 9.488 (computed at a significance level of 0.05 and 4 degrees of freedom), which means if the ${\chi}^2$ statistic is greater than 9.488 the null hypothesis can be rejected at 5\% significance level}
\label{table:chi}
\end{table*}

\begin{table*}[h!]
\small
\centering
\begin{tabular}{l | c c | c c | c c}
\toprule

& \multicolumn{2}{c|}{\textbf{\makecell{Human vs\\GPT-4 (vanilla)}}} &  \multicolumn{2}{c|}{\textbf{\makecell{Human vs\\GPT-4 (empathy-defined)}}} & \multicolumn{2}{c}{\textbf{\makecell{GPT-4 (vanilla) vs \\GPT-4 (empathy-defined)}}} \\

\textbf{} & \textbf{${\chi}^2$-value} & \textbf{p-value} & \textbf{${\chi}^2$-value} & \textbf{p-value} & \textbf{${\chi}^2$-value} & \textbf{p-value} \\

\midrule

All emotions & \textbf{112.136} & \textbf{4.467e-25***} & \textbf{134.124} & \textbf{7.504e-30***} & 1.091 & 0.580 (p > 0.05)\\

Positive & \textbf{90.694} & \textbf{2.023e-20***} & \textbf{92.406} & \textbf{8.595e-21***} & 0.706 & 0.703 (p > 0.05)\\

Negative & \textbf{36.709} & \textbf{1.068e-08***} & \textbf{51.939} & \textbf{5.268e-12***} & 2.136 & 0.344 (p > 0.05)\\

\bottomrule

\end{tabular}
\caption{Statistical ${\chi}^2$ test results corresponding to the proportions of \textit{Bad}, \textit{Okay}, and \textit{Good} empathy ratings of the human's and GPT-4's responses (based on the two prompts). In this case, we compare two by two. The critical value of the ${\chi}^2$ distribution in this case is 5.991 (computed at a significance level of 0.05 and 2 degrees of freedom), which means if the ${\chi}^2$ statistic is greater than 5.991 the null hypothesis can be rejected at 5\% significance level}
\label{table:chi_2}
\end{table*}

\section{Change of Average Empathy Ratings with Evaluator's Empathy Propensity}
\label{app:empathy_propensity}

\begin{figure*}
     \centering     
     \subfloat[\centering Human Responses]{{\includegraphics[width=0.33\textwidth]{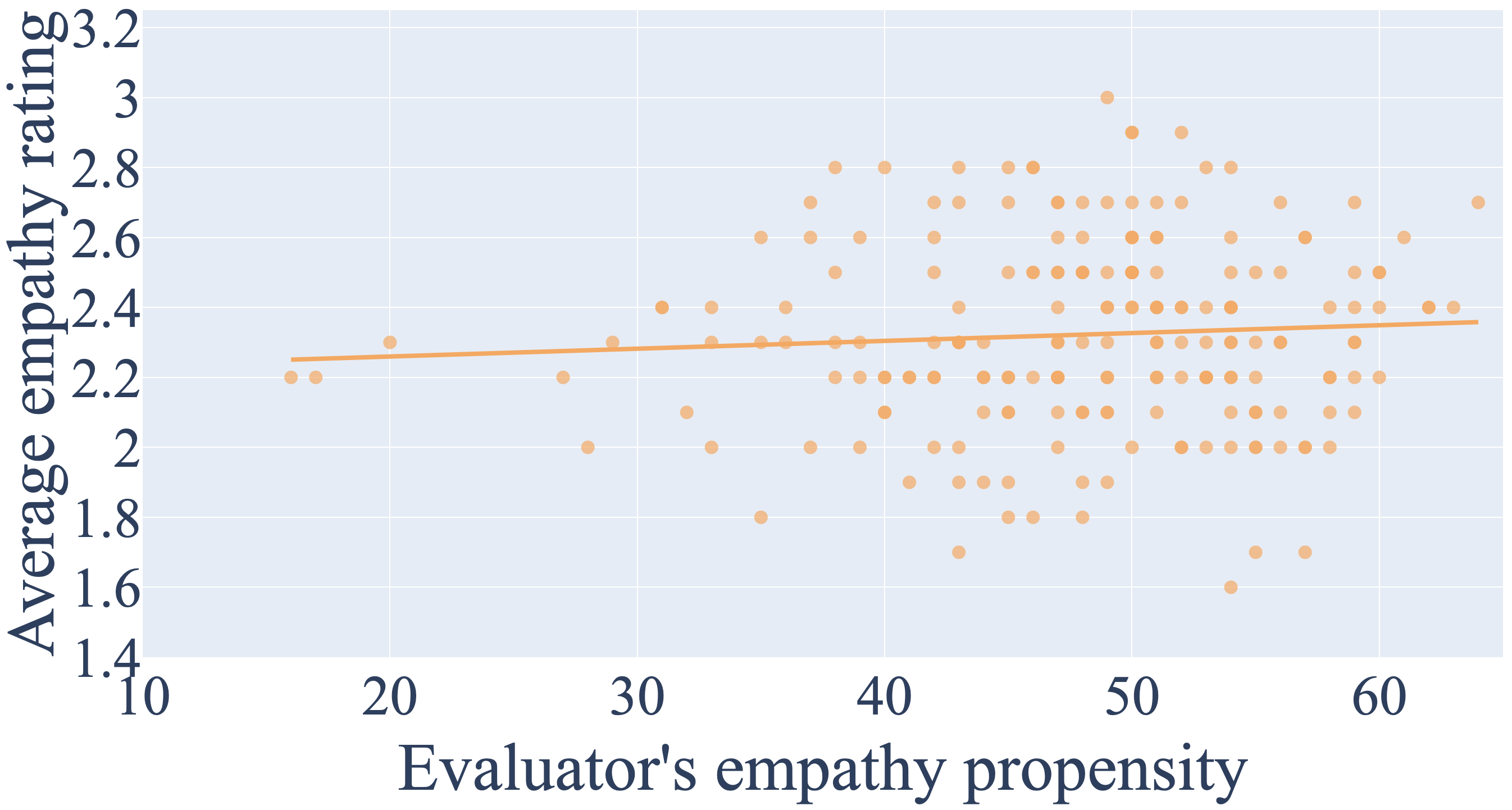}}}     
     \subfloat[\centering GPT-4 (Vanilla) Responses]{{\includegraphics[width=0.33\textwidth]{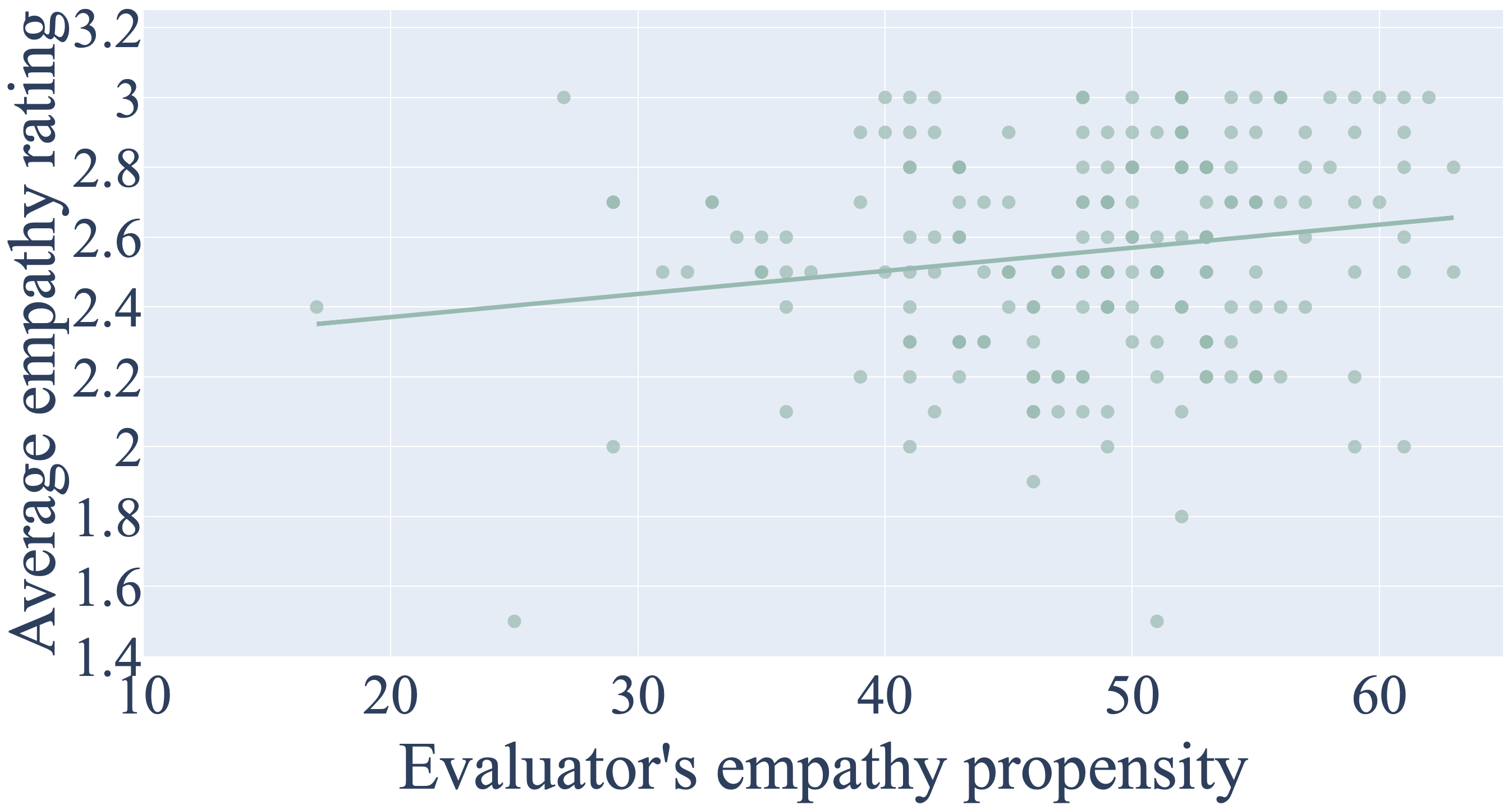}}}
     \subfloat[GPT-4 (Empathy-defined) Responses]{{\includegraphics[width=0.33\textwidth]{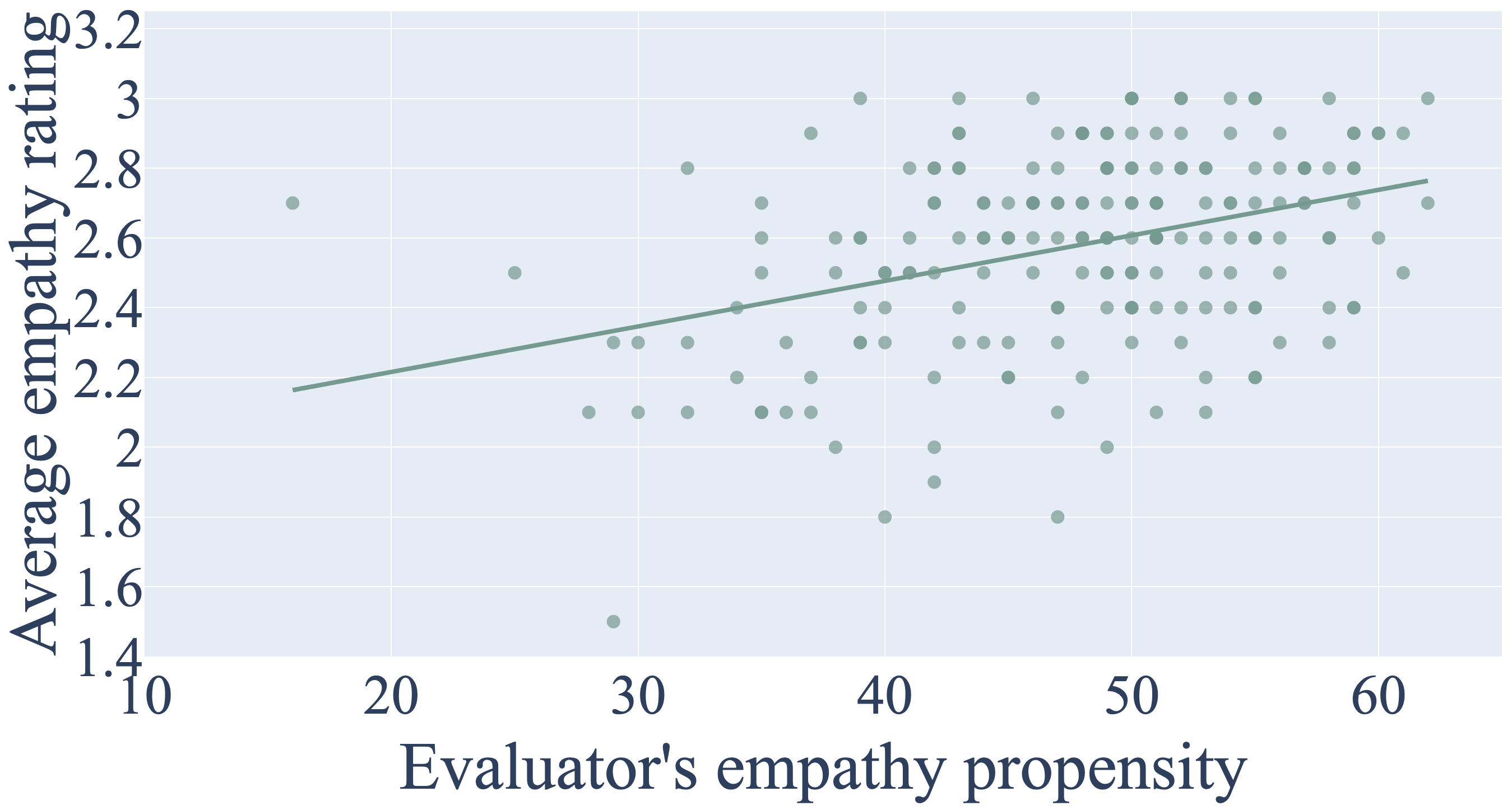}}}  
     \caption{The scatter plots of average empathy ratings against evaluator's propensity to empathize as measured by the Toronto Empathy Questionnaire. Ordinary Least Squares (OLS) regression trendlines are also plotted.}
     \label{fig:empathy_propensity}
\end{figure*}

Figure \ref{fig:empathy_propensity} shows scatter plots corresponding to the evaluators' average empathy ratings of the responses against their empathy propensities as measured by the Toronto Empathy Questionnaire. Regression trendlines based on Ordinary Least Squares (OLS) regression are also shown for each of humans' and the two versions of GPT-4's response ratings. The t-test statistics corresponding to the slopes of the trendlines are denoted in Table \ref{table:propensity}. 

\begin{table*}[ht!]
\small
\centering
\begin{tabular}{l | c c c c}
\toprule

& \textbf{slope} & \textbf{std err} & \textbf{t-value} & \textbf{p-value} \\

\midrule

Human & 0.0022 & 0.002 & 0.967 & 0.335\\
GPT-4 (vanilla) & 0.0066 & 0.003 & \textbf{2.461} & \textbf{0.015*}\\
GPT-4 (empathy-defined) & 0.0130 & 0.002 & \textbf{5.653} & \textbf{0.000***} \\

\midrule

\multicolumn{5}{l}{\textbf{t-test statistics for the difference in 2 independent slopes}\vspace{0.5mm}} \\

& \multicolumn{2}{c}{\textbf{t-value}} & \multicolumn{2}{c}{\textbf{p-value}} \\

Human vs GPT-4 (vanilla) & \multicolumn{2}{c}{1.215} & \multicolumn{2}{c}{0.225 (p > 0.05)} \\

Human vs GPT-4 (empathy-defined) & \multicolumn{2}{c}{\textbf{3.822}} & \multicolumn{2}{c}{\textbf{0.00015***}} \\

GPT-4 (vanilla) vs GPT-4 (empathy-defined) & \multicolumn{2}{c}{1.783} & \multicolumn{2}{c}{0.075 (p > 0.05)} \\

\bottomrule

\end{tabular}
\caption{The statistics of the Ordinary Least Squares (OLS) regression trendlines of average empathy ratings against evaluator's propensity to empathize (as measured by the Toronto Empathy Questionnaire). t>2 in GPT-4 (vanilla) and GPT-4 (empathy-defined) indicate that the slope of the trendline is statistically significantly different from zero. t>2 for the difference in the slopes of the average empathy ratings of the responses of the human and GPT-4 (empathy-defined) indicates that there is a statistically significant difference between the two slopes.}
\label{table:propensity}
\end{table*}

\section{Participants' Demographics}
\label{app:demographics}

\begin{figure*}[ht!]
     \centering     
     \subfloat[\centering Human]{{\includegraphics[width=0.3125\textwidth]{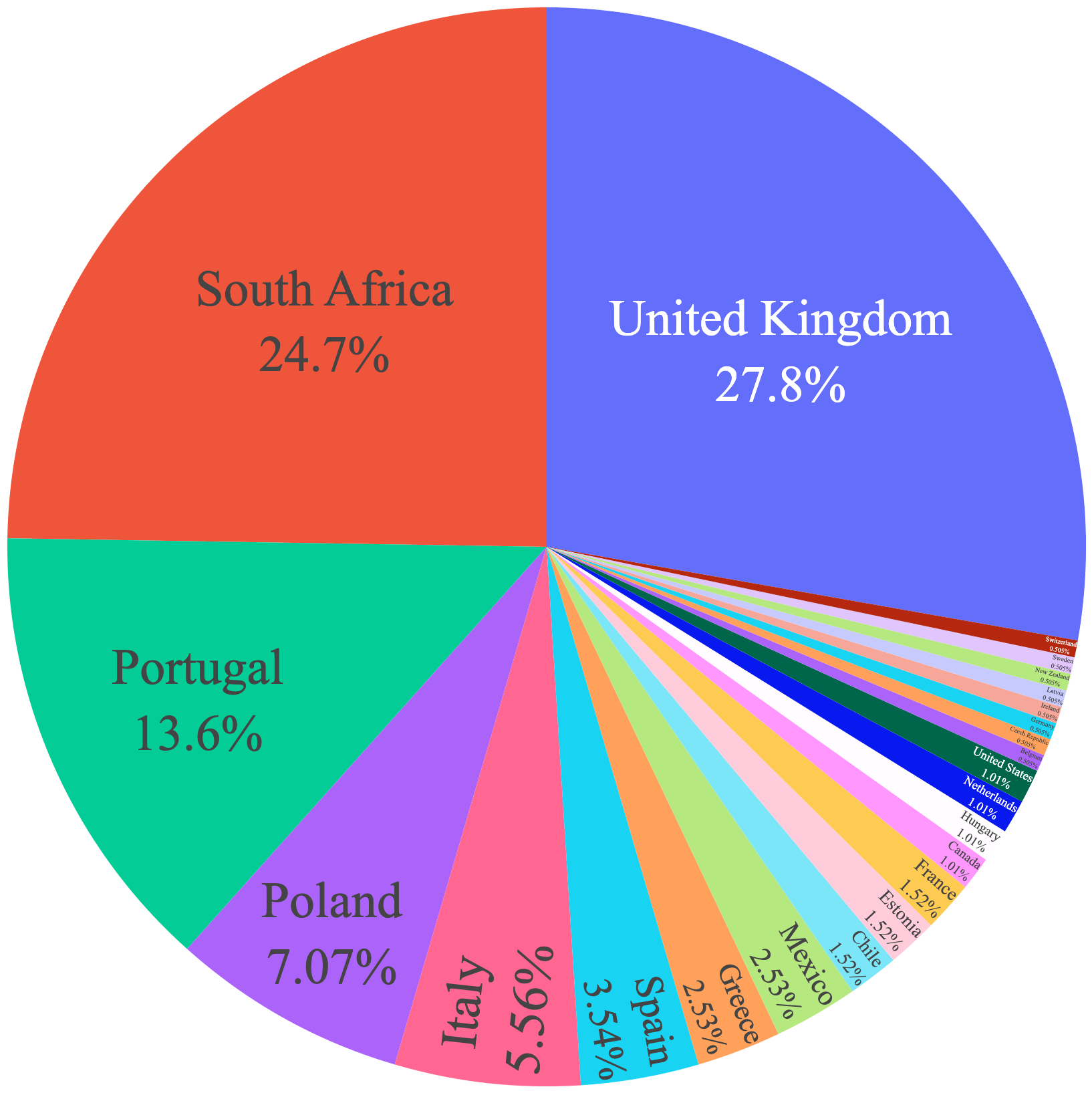}}}     
     \subfloat[\centering GPT-4 (vanilla)]{{\includegraphics[width=0.3125\textwidth]{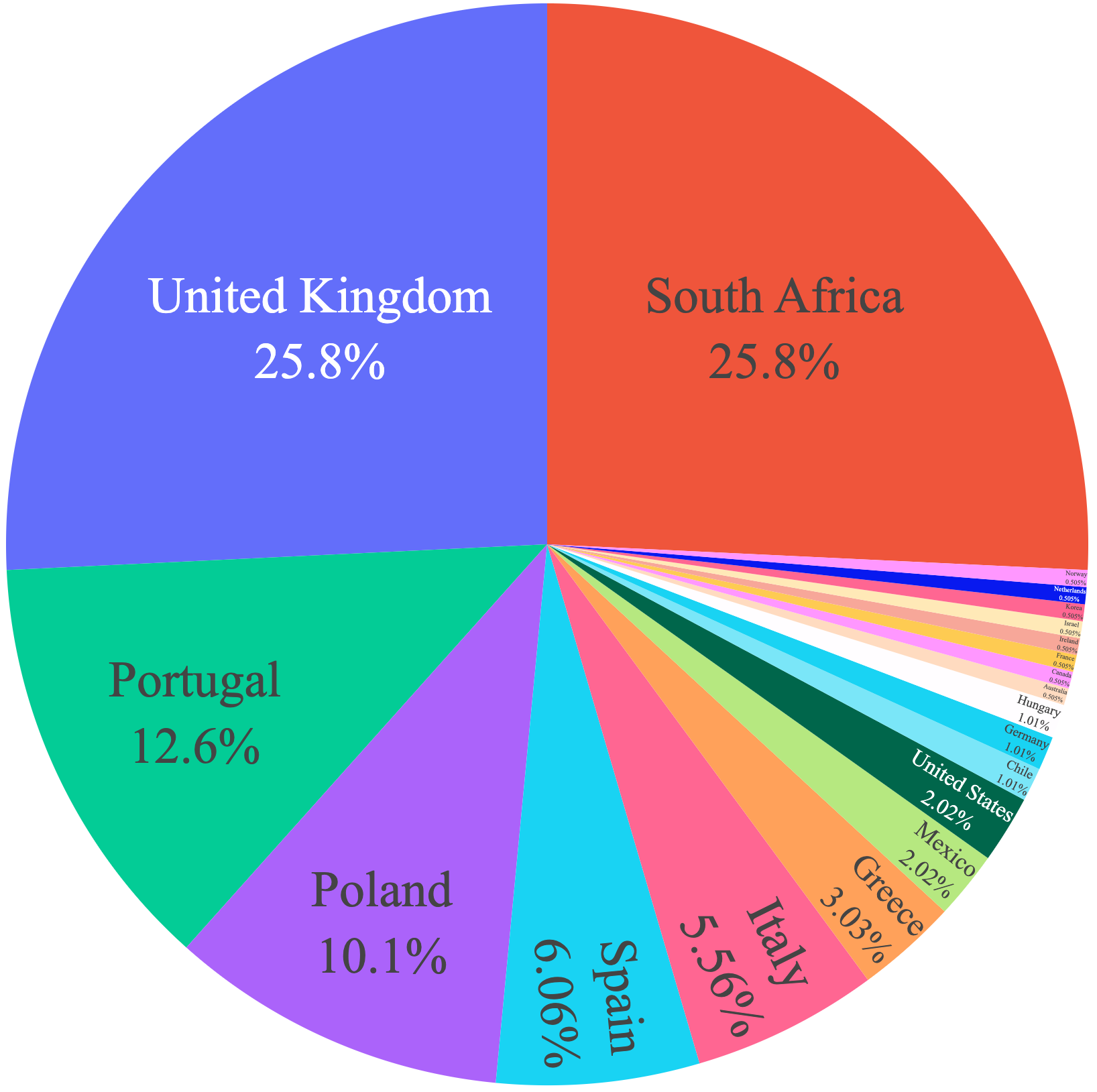}}}
     \subfloat[\centering GPT-4 (empathy-defined)]{{\includegraphics[width=0.375\textwidth]{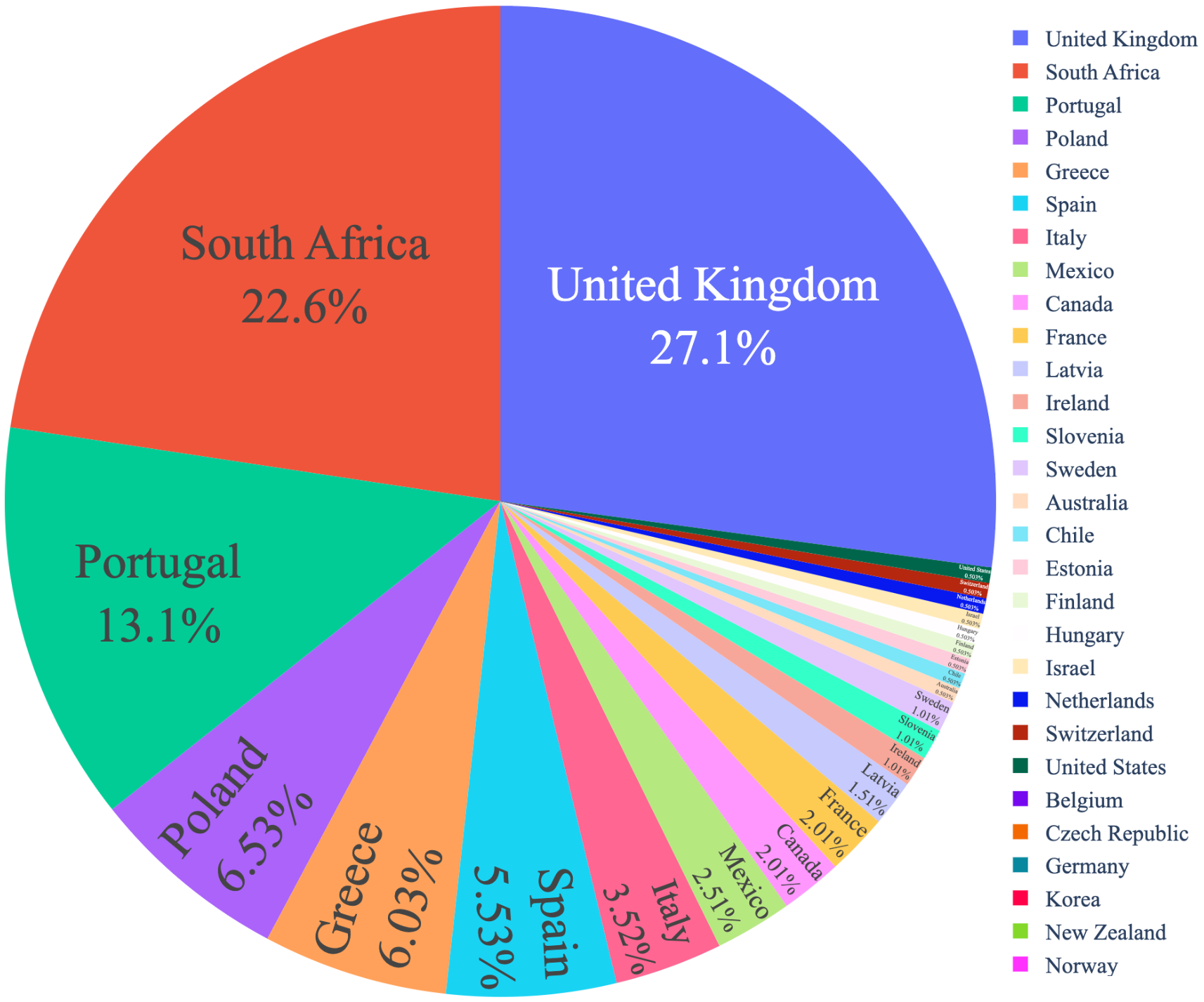}}}     
     \caption{Distribution of the countries of residence of the participants across the three groups.}
     \label{fig:country}
\end{figure*}

\begin{figure*}[ht!]
     \centering     
     \subfloat[\centering Human]{{\includegraphics[width=0.285\textwidth]{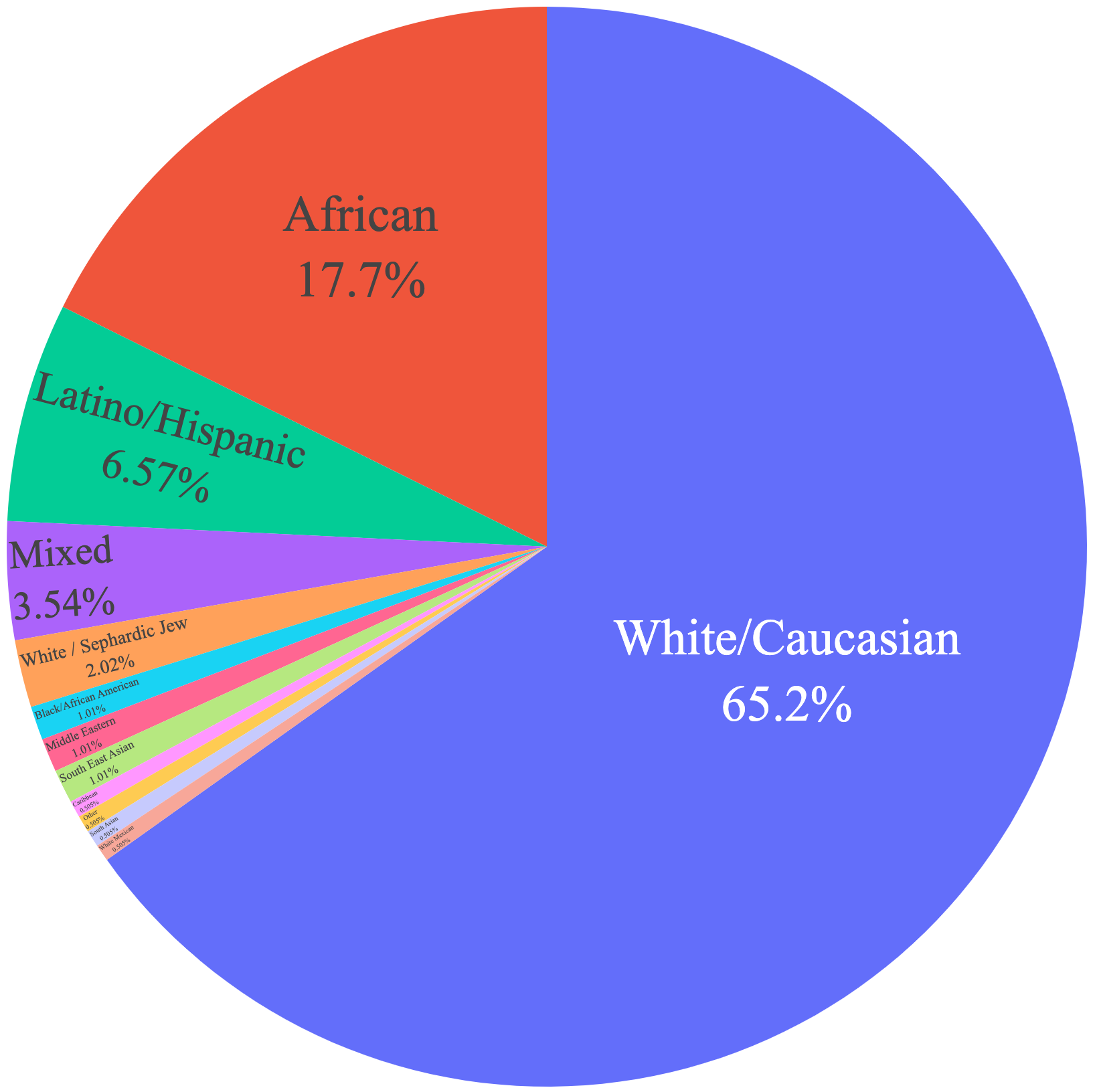}}}     
     \subfloat[\centering GPT-4 (vanilla)]{{\includegraphics[width=0.285\textwidth]{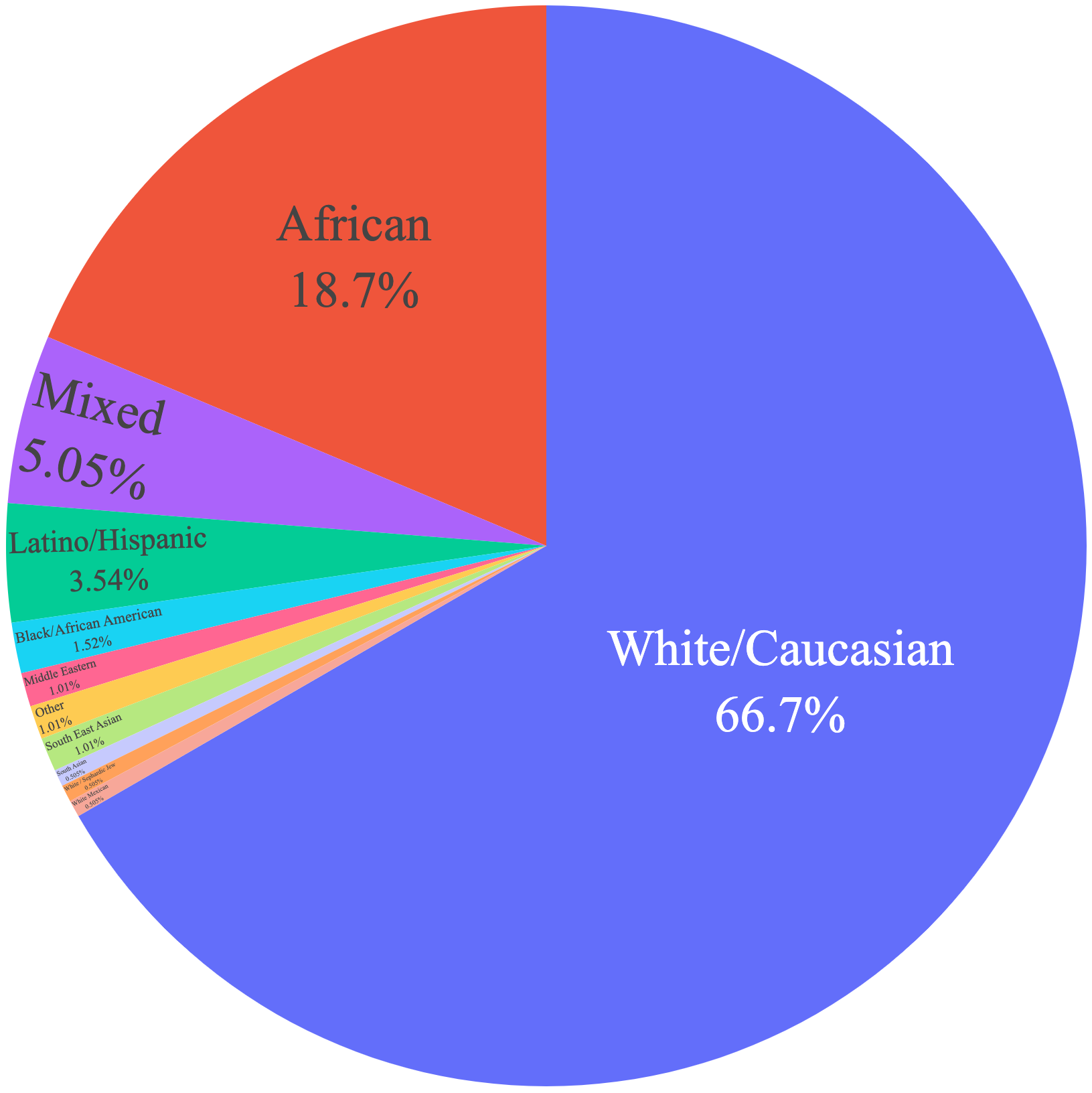}}}
     \subfloat[\centering GPT-4 (empathy-defined)]{{\includegraphics[width=0.4225\textwidth]{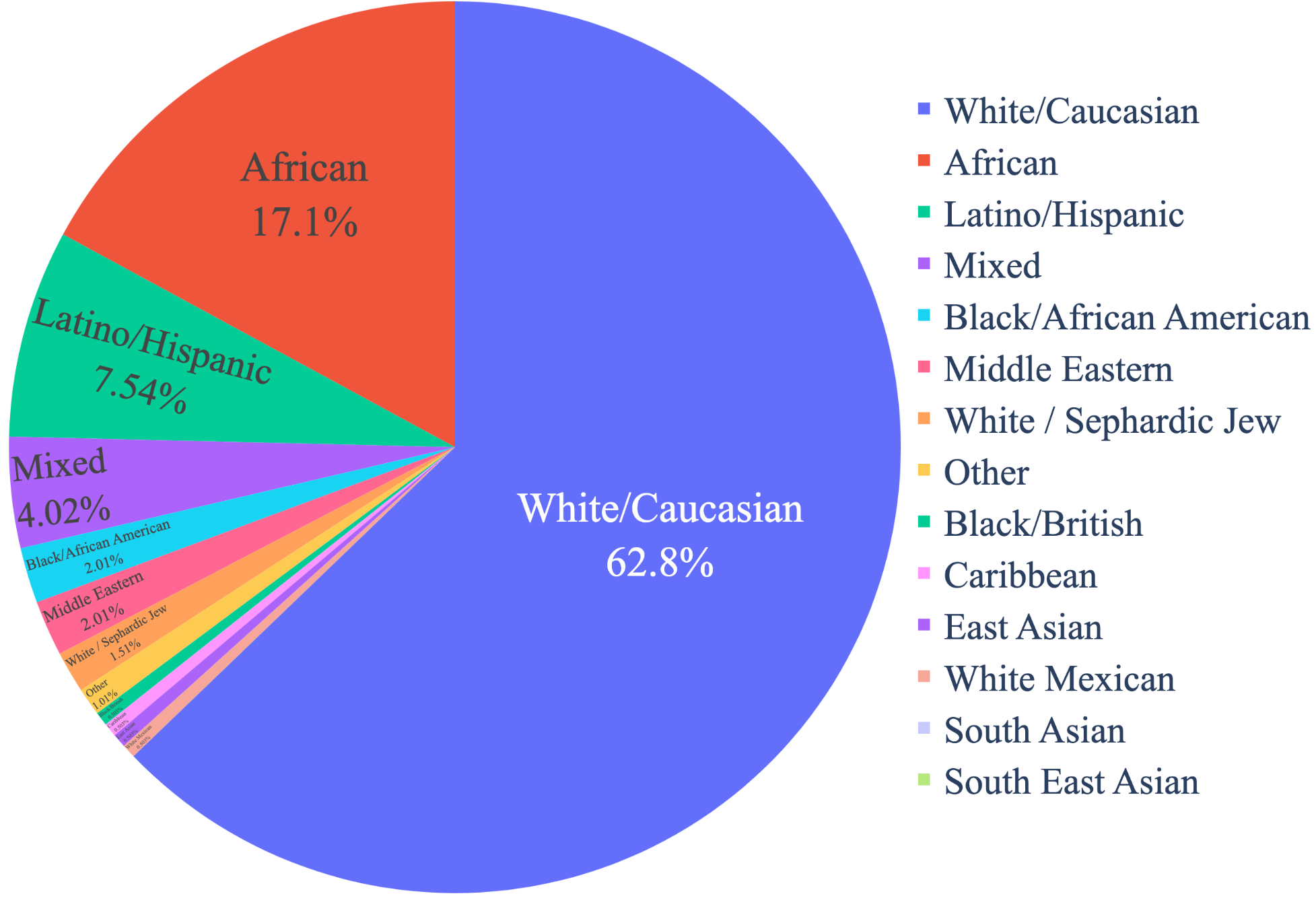}}}     
     \caption{Distribution of the ethnicities of the participants across the three groups.}
    \label{fig:ethnicity}
\end{figure*}

Figures \ref{fig:country} and \ref{fig:ethnicity} respectively show the distributions of the countries of residence and the ethnicities of the participants who rated the three groups of responses. It could be observed that though there are imbalances across the countries and the ethnicities represented in the participants' pool, these demographics are similar across the three groups of participants. This allows control for factors other than the independent variable influencing the results of the study and fair comparison of response ratings across the three groups.

\section{Distribution of Participants' Empathy Propensities}
\label{app:propensity}

\begin{figure*}
     \centering     
     \subfloat[\centering Human response raters]{{\includegraphics[width=0.33\textwidth]{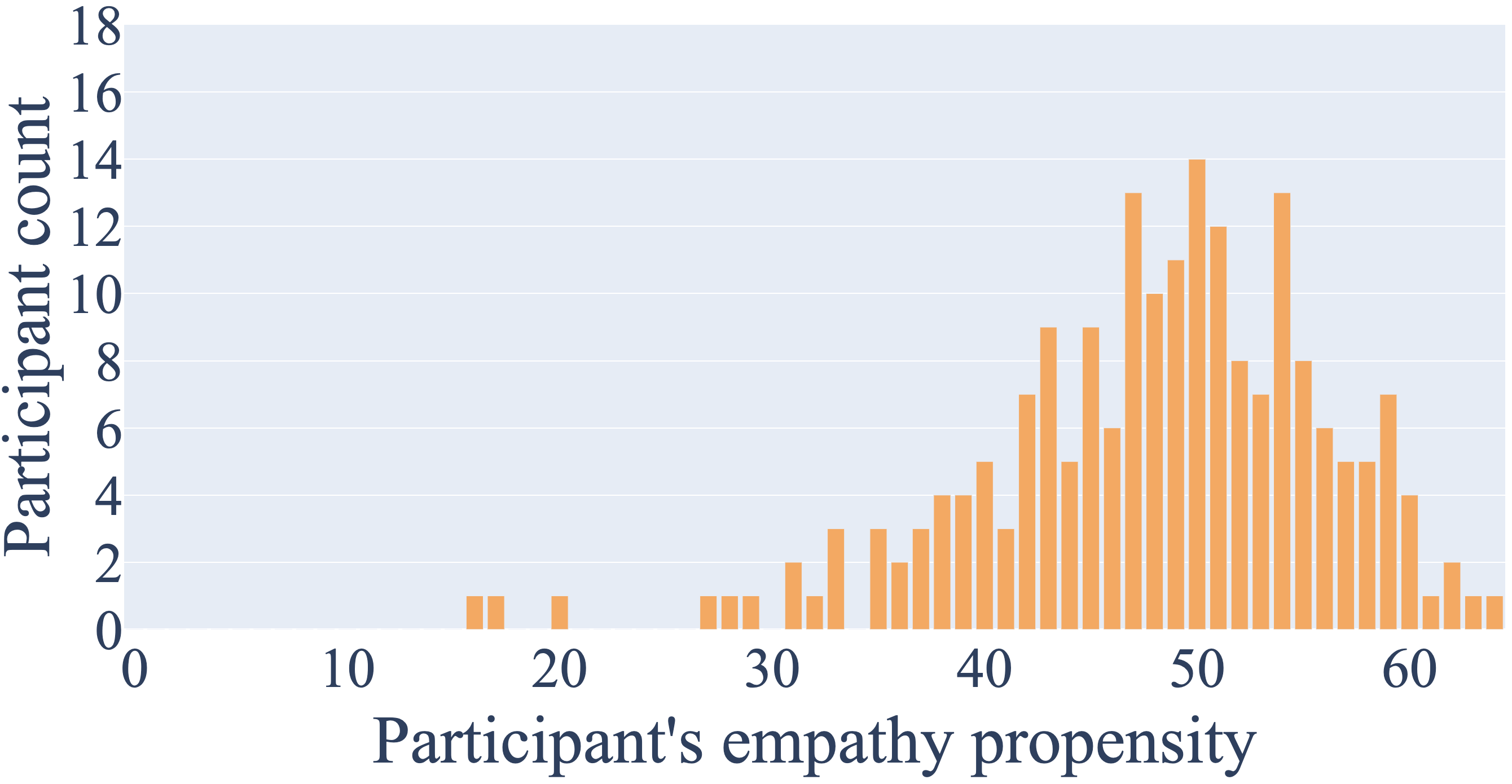}}}     
     \subfloat[\centering GPT-4 (vanilla) responses raters]{{\includegraphics[width=0.33\textwidth]{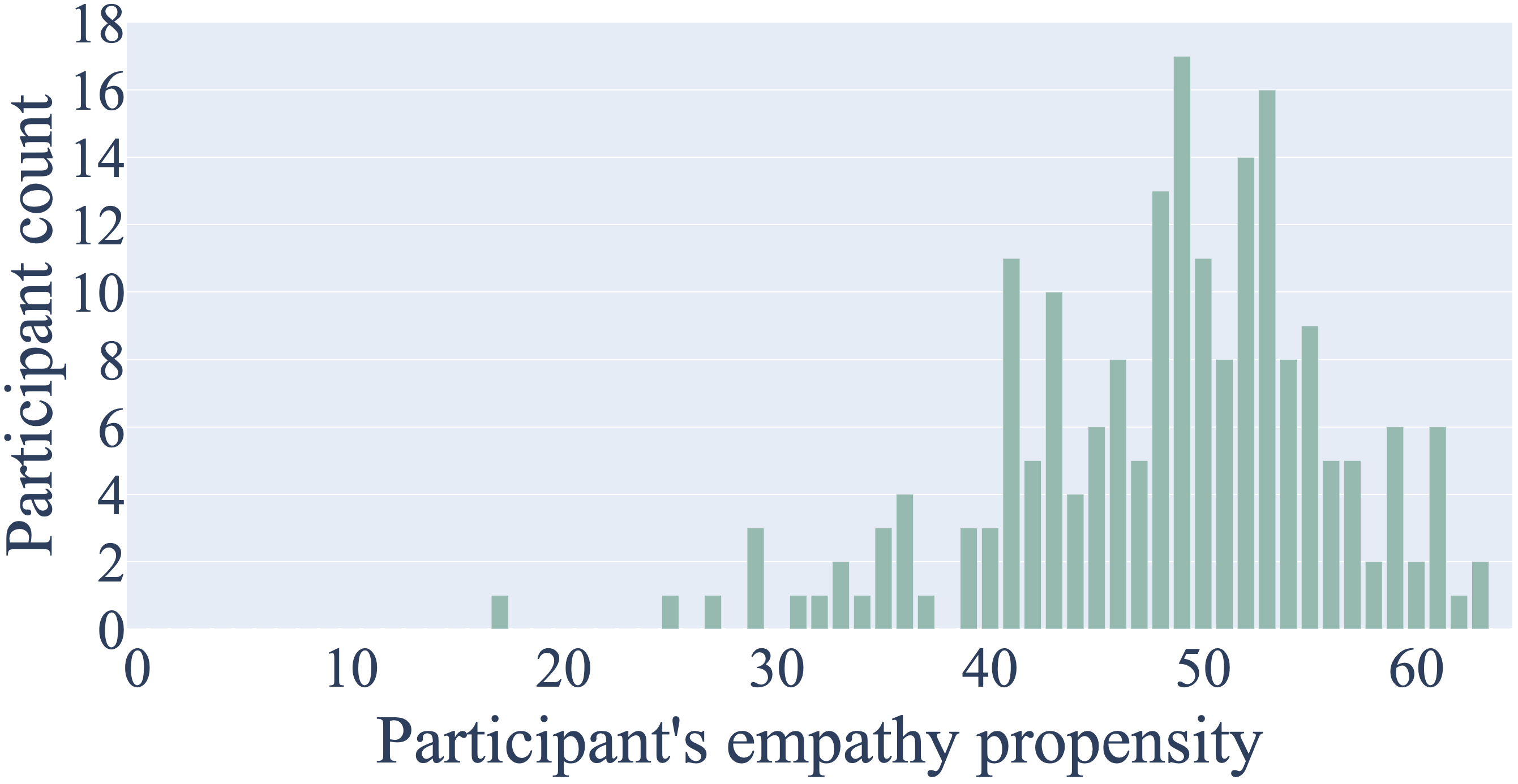}}}
     \subfloat[GPT-4 (empathy-defined) responses raters]{{\includegraphics[width=0.33\textwidth]{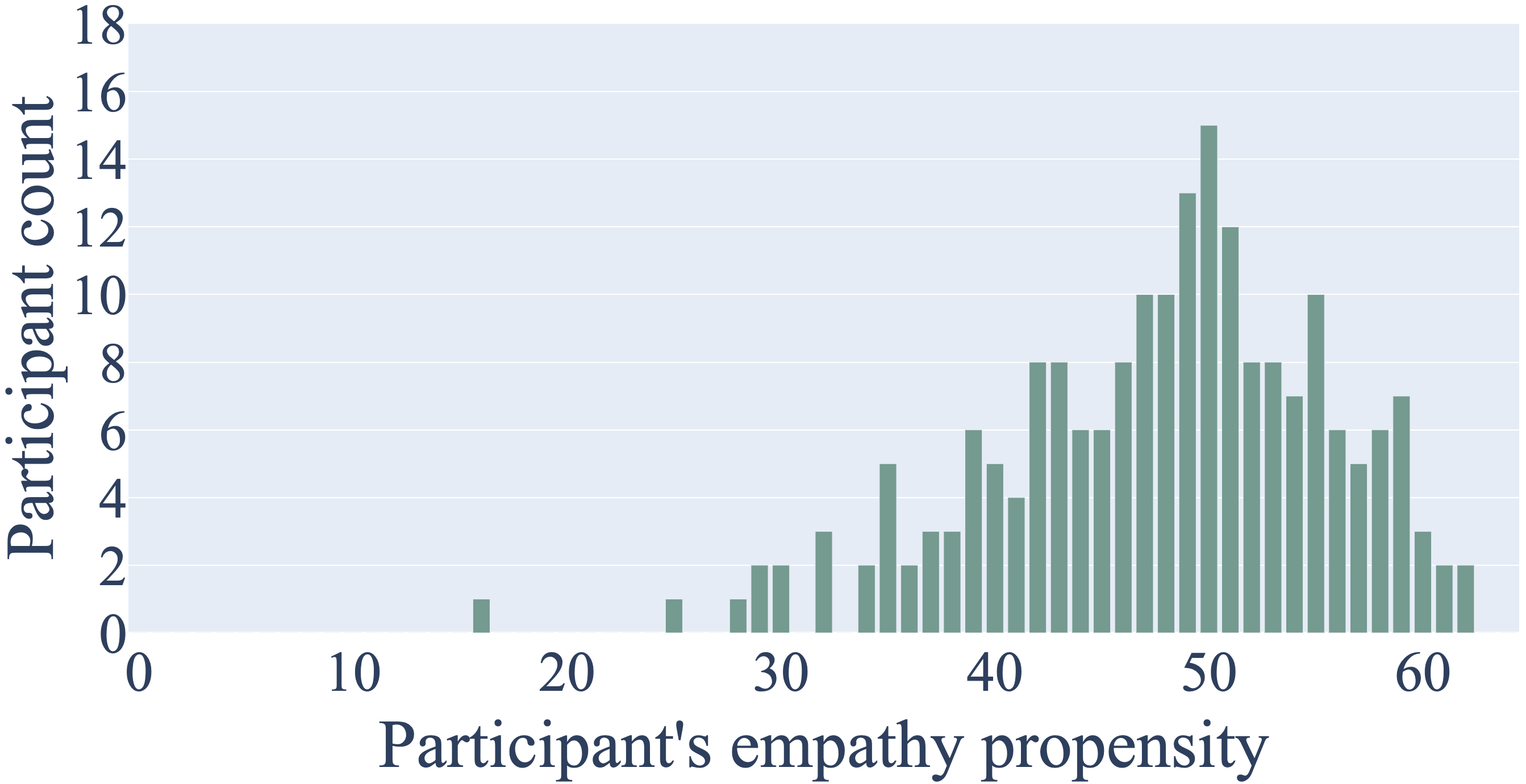}}}     
     \caption{The distributions of the participants' propensities to empathize across the three groups.}
     \label{fig:propensity}
\end{figure*}

Figure \ref{fig:propensity} shows the distributions of the participants' propensities to empathize across the three groups. It could be observed that they are more or less equally distributed across the three groups avoiding any biases in the results that might be caused by any inequal distribution of empathy propensities across the three groups. 

\section{Quality Analysis}
\label{app:quality}

\begin{figure}[H]
  \includegraphics[width=\linewidth]{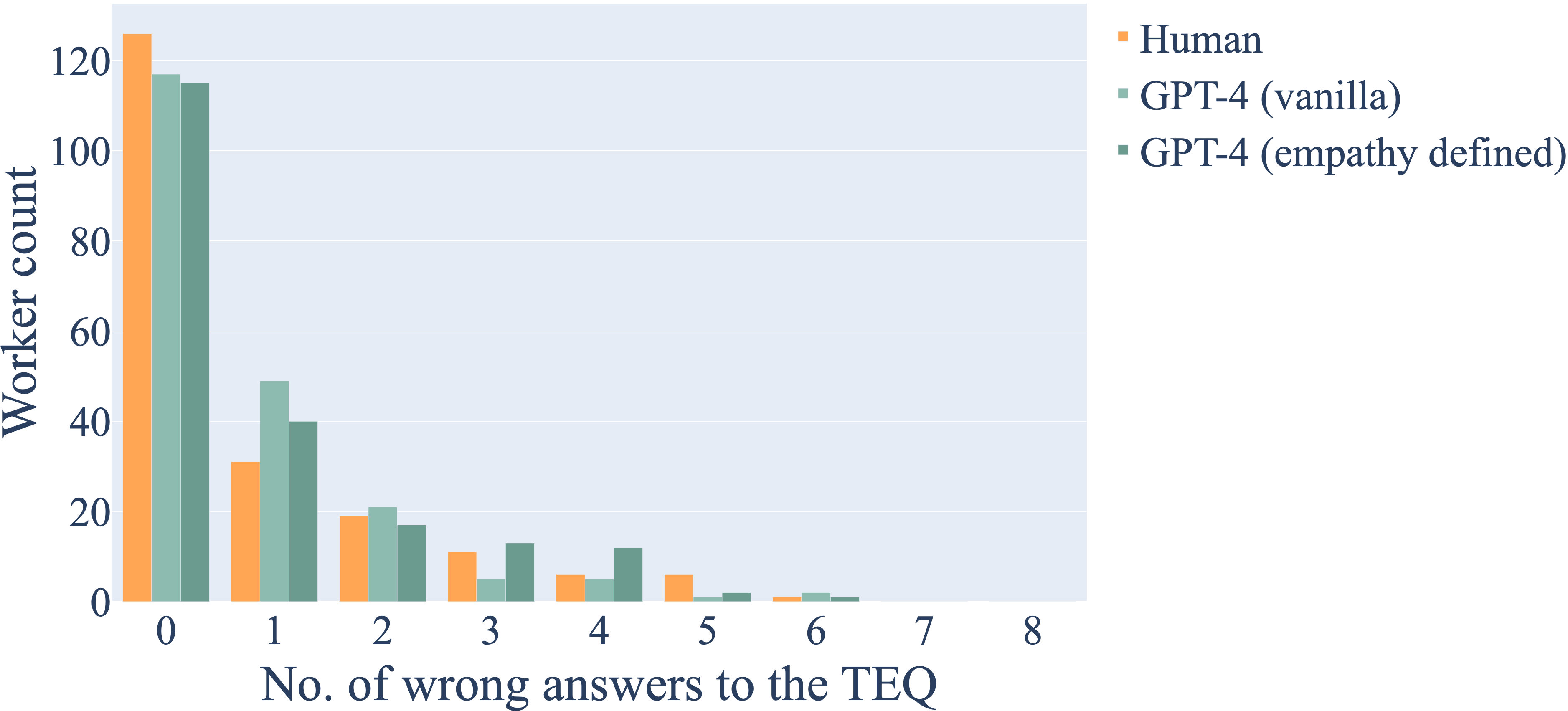}
  \caption{The number of reverse scale questions in the TEQ that were marked wrong by the participants rating the three response groups.}
  \label{fig:teq_histo}
\end{figure}

Figure \ref{fig:teq_histo} shows the number of reverse scale questions in the TEQ that were marked incorrect by the participants rating the three response groups. It was observed that $\approx$60\% of all participants did not get any reverse scale questions wrong and only 2.17\% of all participants got more than half of the reverse scale questions wrong. These statistics validate the quality of the workers recruited for the study. 

\begin{figure}[H]
  \includegraphics[width=\linewidth]{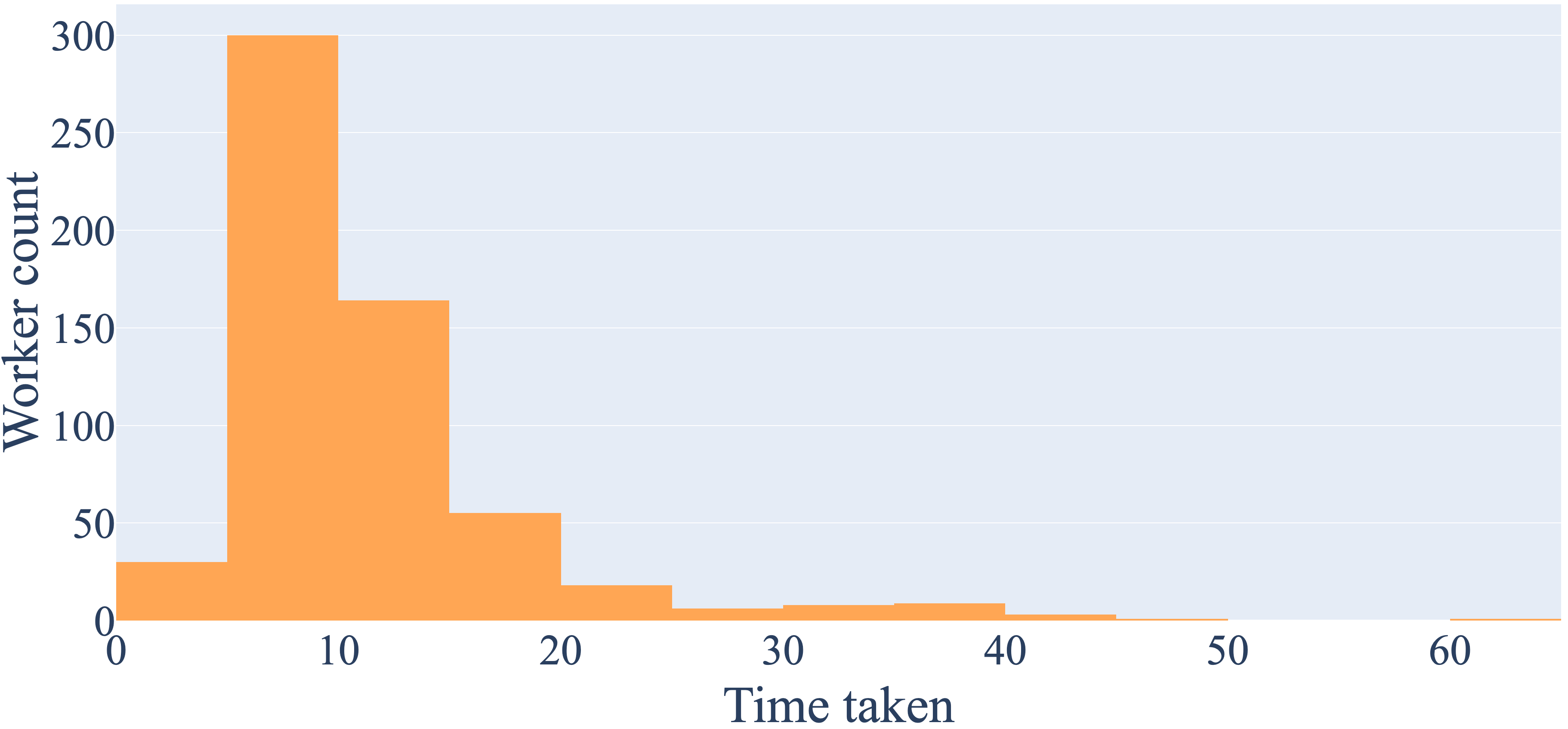}
  \caption{The histogram of times taken to complete the task by all participants.}
  \label{fig:time_histo}
\end{figure}

Further, Figure \ref{fig:time_histo} shows the histogram of times (in minutes) taken to complete the study. On average it took 11 minutes and 18 seconds to complete rating 10 responses, which was close to the average completion time of 15 minutes that we estimated before conducting the study. Only 5\% of all participants were observed to take less than 5 minutes to complete the study, which indicates that most of the participants took time to carefully read the instructions and respond to the questions attentively. 


\section{MITI Classification}
\label{app:miti}

\begin{figure}
     \centering     
     \subfloat[\centering Human responses]{{\includegraphics[width=\linewidth]{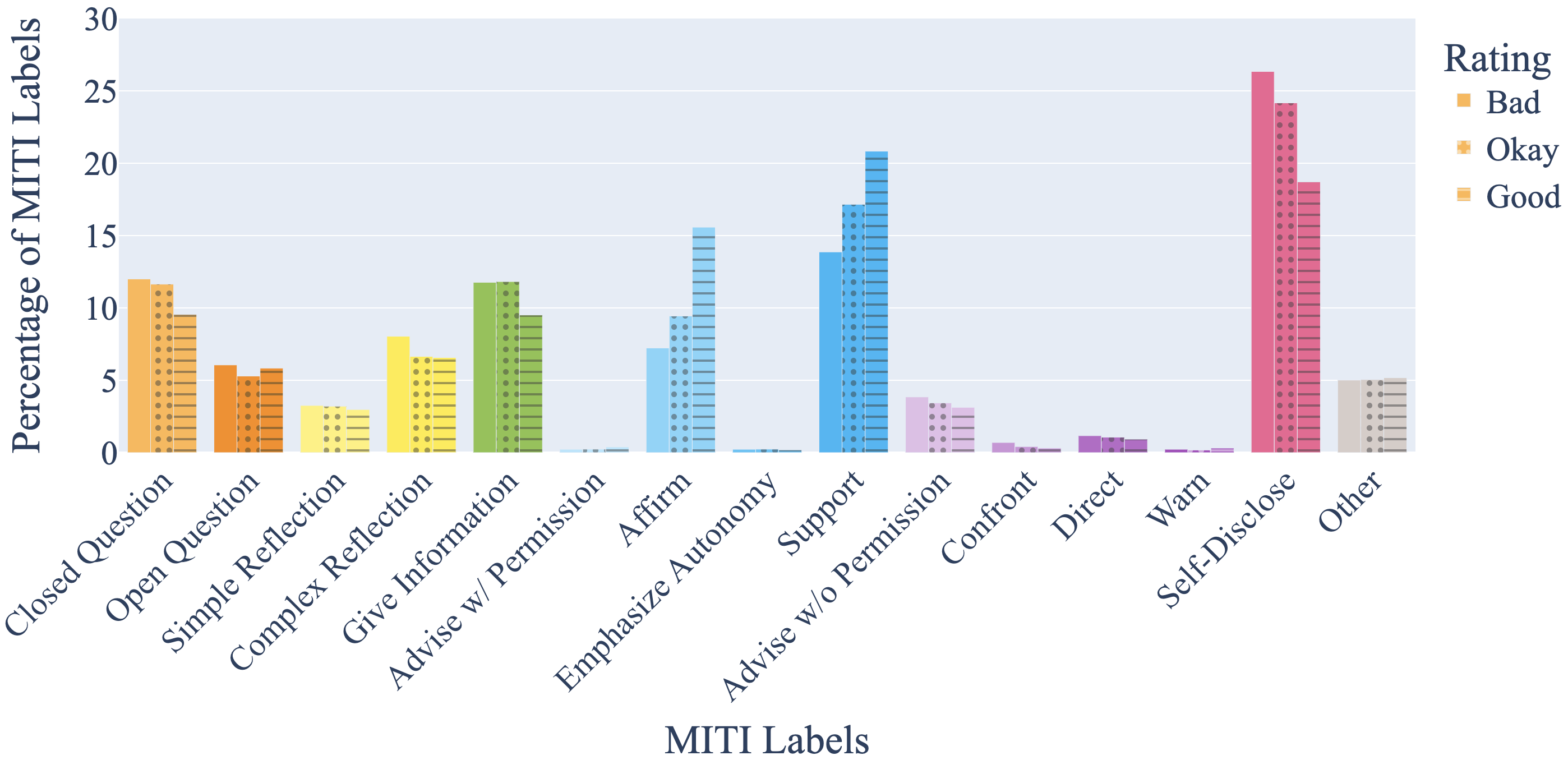}}}     
     
     \subfloat[\centering GPT-4 (vanilla) responses]{{\includegraphics[width=\linewidth]{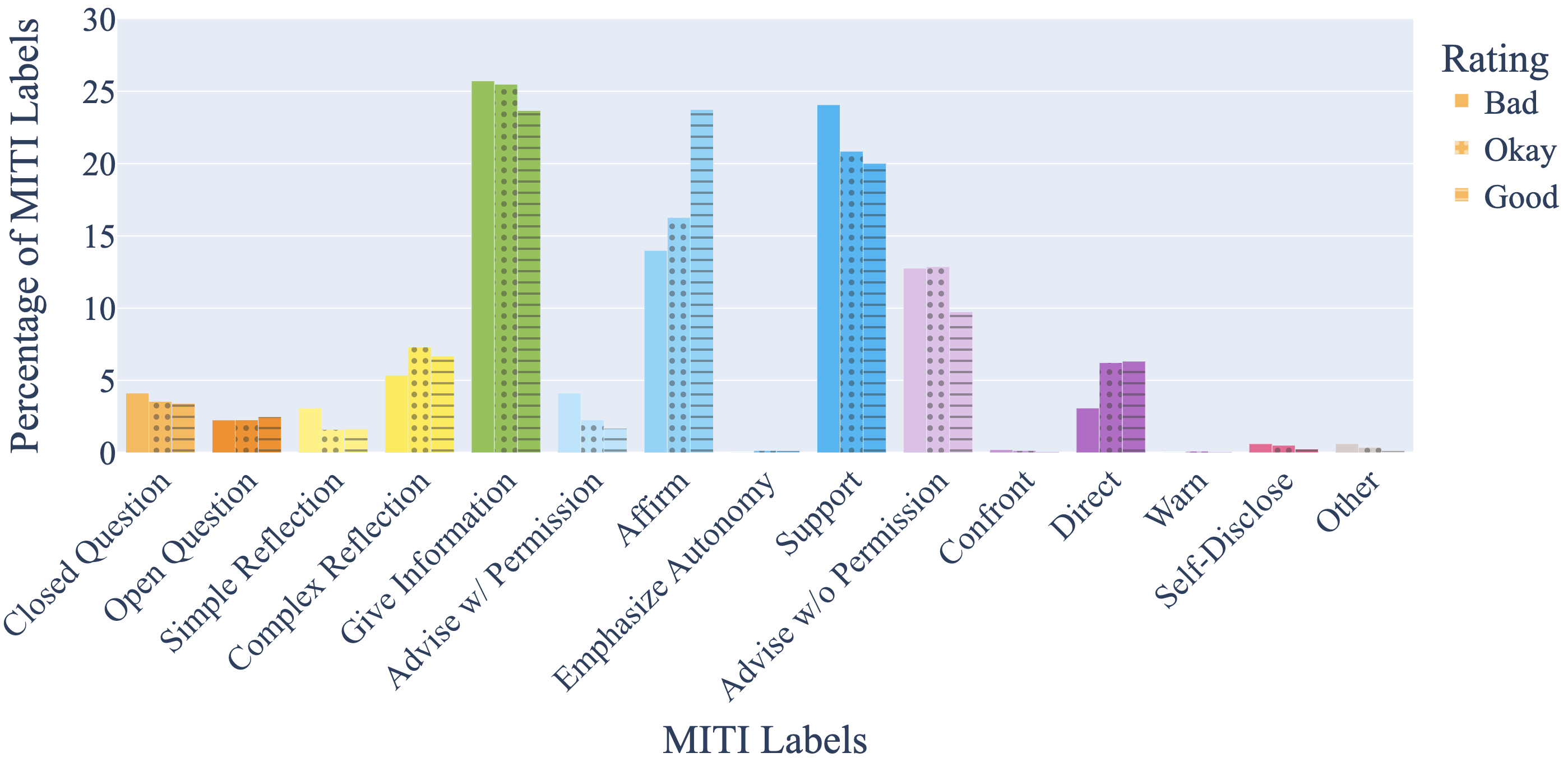}}}
     
     \subfloat[\centering GPT-4 (empathy-defined) responses]{{\includegraphics[width=\linewidth]{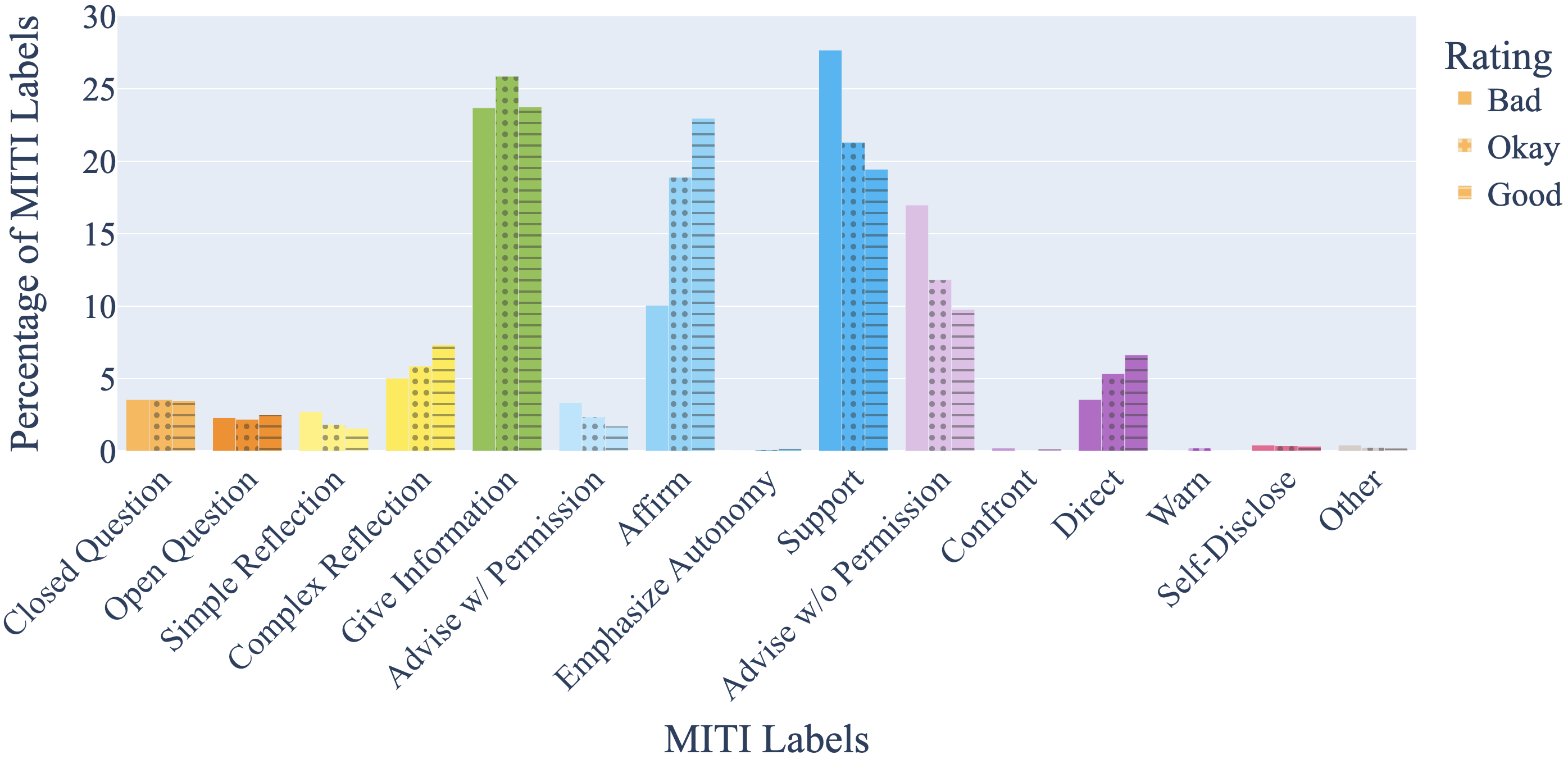}}} 
     
     \caption{The distributions of the percentages of MITI labels present in responses generated by humans and the two versions of GPT-4 that were rated \textit{Bad}, \textit{Okay}, and \textit{Good}.}  
     \label{fig:miti}
\end{figure}

Table \ref{tab:miti} shows the percentage of labels from the Motivational Interviewing Treatment Integrity (MITI) code \cite{miti_4_2_1} present in responses generated by humans and the two versions of GPT-4 that were rated \textit{Bad}, \textit{Okay}, and \textit{Good}, as classified by the classifier introduced by Welivita and Pu \shortcite{boosting}. Figure \ref{fig:miti} visualizes these distributions of the percentages of labels for responses generated by humans and the two versions of GPT-4, separately.

\begin{table*}[ht!]
\small
\centering
\begin{tabular}{l | r r r | r r r | r r r}
\toprule

\multirow{2}{*}{\textbf{MITI label}} & \multicolumn{3}{c}{\textbf{Human}} & \multicolumn{3}{c}{\textbf{GPT-4 (vanilla)}} & \multicolumn{3}{c}{\textbf{GPT-4 (empathy-defined)}}   \\

 & \textbf{\textit{Bad}} & \textbf{\textit{Okay}} & \textbf{\textit{Good}} & \textbf{\textit{Bad}} & \textbf{\textit{Okay}} & \textbf{\textit{Good}} & \textbf{\textit{Bad}} & \textbf{\textit{Okay}} & \textbf{\textit{Good}} \\

\midrule

\multicolumn{10}{l}{\textbf{MI Adherent Response Types:}\vspace{1mm}}\\

Closed Question  &  \textbf{12.00}  &  11.64  &  9.55  &  \textbf{4.12}  &  3.55  &  3.41  &  \textbf{3.56}  &  \textbf{3.56}  &  3.47 \\
Open Question  &  \textbf{6.06}  &  5.29  &  5.84  &  2.26  &  2.27  &  \textbf{2.48}  &  2.31  &  2.20  &  \textbf{2.50} \\
Simple Reflection  &  \textbf{3.26}  &  3.21  &  2.98  &  \textbf{3.09}  &  1.60  &  1.64  &  \textbf{2.73}  &  1.83  &  1.58 \\
Complex Reflection  &  \textbf{8.04}  &  6.65  &  6.57  &  5.35  &  \textbf{7.31}  &  6.68  &  5.03  &  5.86  &  \textbf{7.34} \\
Give Information  &  11.77  &  \textbf{11.82}  &  9.51  &  \textbf{25.72}  &  25.49  &  23.67  &  23.69  &  \textbf{25.86}  &  23.75 \\
Advise with Permission  &  0.23  &  0.24  &  \textbf{0.39}  &  \textbf{4.12}  &  2.27  &  1.68  &  \textbf{3.35}  &  2.36  &  1.73 \\
Affirm  &  7.23  &  9.44  &  \textbf{15.58}  &  13.99  &  16.27  &  \textbf{23.74}  &  10.06  &  18.90  &  \textbf{22.96} \\
Emphasize Autonomy  &  0.23  &  \textbf{0.24}  &  0.19  &  0.00  &  \textbf{0.15}  &  0.14  &  0.00  &  0.10 & \textbf{0.16} \\
Support  &  13.87  &  17.16  &  \textbf{20.83}  &  \textbf{24.07}  &  20.85  &  20.01  &  \textbf{27.67}  &  21.31  &  19.45\vspace{1mm} \\

\multicolumn{10}{l}{\textbf{MI Non-adherent Response Types:}\vspace{1mm}}\\

Advise without Permission  &  \textbf{3.85}  &  3.44  &  3.13  &  \textbf{12.76}  &  12.87  &  9.75  &  \textbf{16.98}  &  11.83  &  9.77 \\
Confront  &  \textbf{0.70}  &  0.42  &  0.31  &  \textbf{0.21}  &  0.15  &  0.07  &  \textbf{0.21}  &  0.00  &  0.14 \\
Direct  &  \textbf{1.17}  &  1.07  &  0.93  &  3.09  &  6.23  &  \textbf{6.32}  &  3.56  &  5.34  &  \textbf{6.64} \\
Warn  &  0.23  &  0.18  &  \textbf{0.31}  &  0.00  & \textbf{0.10}  &  0.05  &  0.00  &  \textbf{0.21}  &  0.00\vspace{1mm}\\

\multicolumn{10}{l}{\textbf{Other:}\vspace{1mm}}\\

Self-Disclose  &  \textbf{26.34}  &  24.17  &  18.71  &  \textbf{0.62}  &  0.51  &  0.25  &  \textbf{0.42}  &  0.37  &  0.34 \\
Other  &  5.01  &  5.05  &  \textbf{5.18}  &  \textbf{0.62}  &  0.36  &  0.14  &  \textbf{0.42}  &  0.26  &  0.20 \\

\bottomrule
\end{tabular}
\caption{The percentages of labels from the MITI code present in responses generated by humans and the two versions of GPT-4 that were rated \textit{Bad}, \textit{Okay}, and \textit{Good}.}
\label{tab:miti}
\end{table*}

\end{document}